\documentclass[floatfix,footinbib,reprint,twocolumn,prb,aps,a4paper,preprintnumbers,amsmath,amssymb,longbibliography]{revtex4-1}
\usepackage{latexsym,amsmath,graphics,graphicx,exscale,colordvi,color,mathtools}
 \pdfoutput=1
 \usepackage{appendix}
\usepackage{soul,xcolor,braket}

\usepackage{multirow}

\DeclareMathOperator{\trace}{Tr}

\begin{document}

\title{Friedel oscillations induced by magnetic skyrmions: from scattering properties to all-electrical detection}
\author{Mohammed Bouhassoune$^{1,2}$}
\author{Samir Lounis$^{1,3}$}\email{s.lounis@fz-juelich.de}
\affiliation{$^1$Peter Gr\"unberg Institut \& Institute for Advanced Simulation, 
Forschungszentrum J\"ulich \& JARA, D-52425 J\"ulich,
Germany\\
$^2$D\'epartement de Physique, FPS, Cadi Ayyad University, 40000 Marrakech, Morocco\\
$^3$Faculty of Physics, University of Duisburg-Essen, 47053 Duisburg, Germany}

\begin{abstract}
Magnetic skyrmions are spin swirling solitonic defects that can play a major role in  information technology. Their future in applications and devices hinges on their efficient manipulation and detection. Here, we explore from ab-initio their nature as magnetic  inhomongeities in an otherwise unperturbed magnetic material, Fe layer covered by a thin Pd film and deposited on top of Ir(111) surface. The presence of skyrmions triggers scattering processes, from which Friedel oscillations emerge. The latter mediate interactions among skyrmions or between skyrmions and other potential surrounding defects. In contrast to their wavelengths, the amplitude of the oscillations depends strongly on the size of the skyrmion.  The analogy with the scattering-off atomic defects enables the assignment of an effective scattering potential and a phase shift to the skyrmionic particles, which can be useful to predict their  behavior on the basis of simple scattering frameworks. The induced charge ripples can be utilized for a noninvasive all-electrical detection of skyrmions located on a surface or even if buried a few nanometers away from the detecting electrode.

\end{abstract}

\maketitle

\section{Introduction}

Materials' hallmarks hinge on the propagation of electrons, which mediate the interaction among atoms and impact nontrivially on   thermal, electrical, optical, magnetic and transport properties. After scattering at localized defects, Friedel oscillations arise, which  lead to charge screening of the foreign atoms, as routinely observed with scanning tunneling spectroscopy (STS) (for example,~\cite{Crommie1993,Hasegawa1993,Crommie1993b,Silly2004,Lounis:2012,Meier2011}). Besides the nature of the impurities,  the electronic states residing on  constant energy contours in the reciprocal space define at a given energy the form, wavelength and amplitude of the produced ripples. The latter can be, for instance, anisotropic or even focused along well defined regions for rather large distances if the energy contours bear flat areas~\cite{Weismann2009,Lounis2011,Prueser2011,Prueser2012,Avotina2006,Bouhassoune2014,Howon2020}. In the context of magnetism, the resulting charge density oscillations govern the magnetic long-range interactions among single atoms, which can drive communication between nanoobjects~\cite{Zhou2010,Khajetoorians2012,Ngo2012,Prueser2014,Khajetoorians2016,Bouaziz2020} and trigger appealing magnetic behavior~\cite{Khajetoorians2012,Ngo2012,Prueser2014,Stepanyuk2007,Brovko2008}. Elastic interactions can emerge giving rise to the  self-assembly of superstructures of adatoms on surfaces~\cite{Silly2004}.

\begin{figure*}[ht!]
	\includegraphics*[angle=0,width=1.\linewidth]{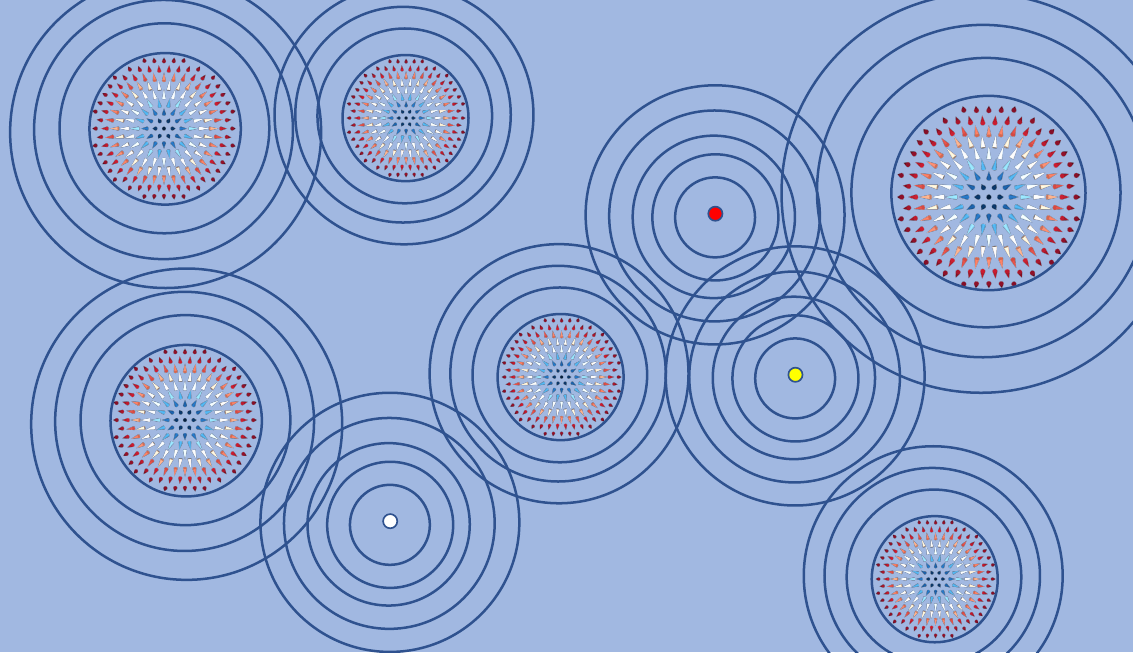}
	\caption{Schematic visualization of Friedel {charge} oscillations {(blue circles)} emanating from conventional impurities {(blue, yellow and white small spheres)} or from skrmionic defects {(noncollinear spin-textures)}. {The interference between the incoming electronic states of the substrate and those scattered away from the various defects induce the charge waves.} Such charge ripples host a wealth of information on the hosting material and on the defects while mediating electronic, magnetic and elastic interactions.}  
	\label{Fig1}
\end{figure*}

Magnetic skyrmions~\cite{Bogdanov,Roessler2006}, i.e., localized noncollinear spin textures of topological nature~\cite{Nagaosa2013} with particle-like properties, are solitonic defects, which {similarly to atomic defects} should scatter the electronic states of the hosting materials. {In Figure~\ref{Fig1}, we represent schematically the Friedel oscillations, shown as blue circles, emerging equally from skyrmions of different sizes or from atomic defects embeded on some surface}. Since skyrmions are heavily prospected as potential magnetic bits for future devices~\cite{Fert2013,Sampaio2013,Tomasello2013,Zhou2014,Crum2015,Zhang2015,Yu2016,Garcia-Sanchez2016,Xia2017,Fernandes2020,Sai2021}, the induced Friedel oscillations could bear interesting information on the nature of the skyrmions and their interaction with the surroundings, which could prove useful for applications. The classical example of a storage device based on skyrmions is a racetrack memory~\cite{Fert2013,Tomasello2013,Parkin2008}, where the magnetic textures would be driven by a current.
Similarly to atomic defects, the emerging charge ripples would govern the distance and the repulsive or attractive  interactions among the sequence of magnetic skyrmions or between the skyrmions and surrounding defects, being solitonic entities, dislocations, edges or simply atomic impurities. {This means that the sequence, density and motion of skyrmions in a device would be controlled by the related Friedel oscillations.} Those {ripples} are the driving mechanism behind the distant-dependent  interactions between magnetic skyrmions observed with Lorentz transmission electron microscopy in B20--FeGe~\cite{Haifeng2018}. {They were} predicted to exist in an Fe monolayer deposited on Pd(111) with an alloyed overlayer made of Pt and Ir~\cite{Rozsa:2016} {and are responsible for the deflection of skyrmions generated in PdFe/Ir surface at a finite distance from atomic defects~\cite{Fernandes2020a,Arjana2020}.} Thus, the understanding and control of the shape, amplitude and wavelength of these interactions, intimately related to the decay of the charge density oscillations, are decisive if the skyrmions are to be used as  building blocks in nanospintronic devices. {This is not limited to the aforementioned racetrack memory but extends to various technological concepts proposed recently~\cite{Sai2021}: ranging from those based on multiple skyrmions for synaptic~\cite{Huang2017,Song2020,Li2017,Chen2018}, reservoir~\cite{Prychynenko2018,Bourianoff2018} and stoachastic~\cite{Pinna2018,Zazvorka2019,Yao2020} devices to  those involving single skyrmions  as those proposed for spin torque nanooscillator based on magnetic tunneling junctions~\cite{Zhang2015b,Finocchio2015}, where interactions with edges and other defects is inevitable. }

Interestingly, model calculations indicated that skyrmion-like  magnetic Friedel  oscillations can be induced by an adatom deposited on a surface characterized by a finite spin-orbit interaction~\cite{Lounis:2012} with a magnetic complexity that is detectable experimentally~\cite{Strozecka:2011}. While free electrons scattering at skyrmions could carry a spin-magnetization of chiral nature~\cite{Denisov:2019}, topological surface states can trigger interesting physics when eventually hitting a skyrmion~\cite{Wang:2020}.

\begin{figure*}[ht!]
\includegraphics*[angle=0,width=\linewidth]{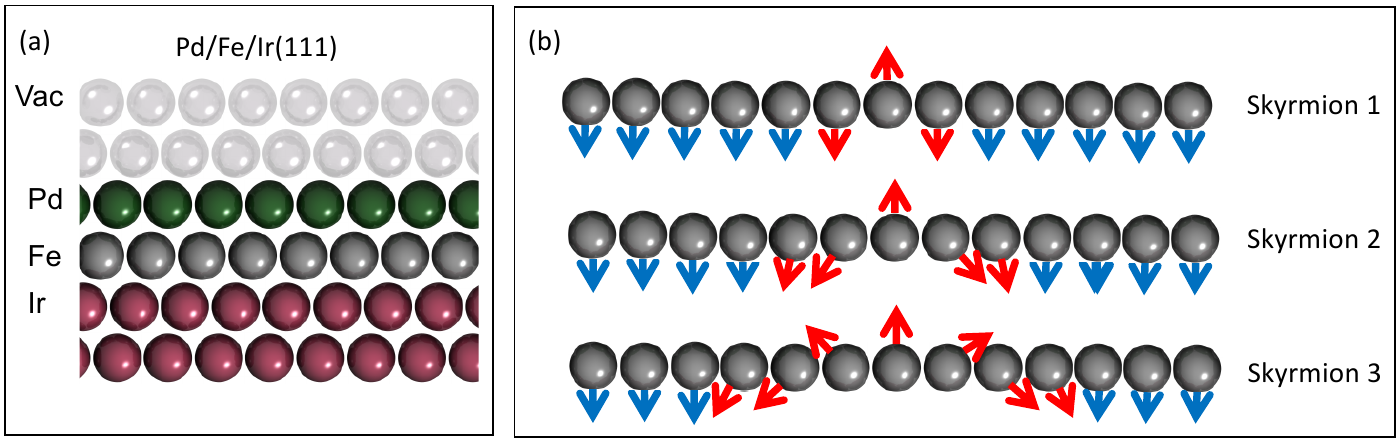}
	\caption{{(\textbf{a}) Illustration of the heterostructure of PdFe bilayer deposited on fcc Ir(111) surface. Vac represents the vacuum  site in which the electronic structure is calculated in order to investigate the Friedel charge oscillations. (\textbf{b}) Side view of the  three confined spin-textures converged in the Fe monolayer: (i) Skyrmion 1, where one of the Fe magnetic moments is flipped with respect to the ferromagnetic background (blue moments), (ii) Skyrmion 2 and (iii) Skyrmion 3 contain respectively 19 and 37 Fe atoms including all their nearest neighbors, which are allowed to have their electronic structure converged~\cite{Crum2015}. }}  
	\label{Fig_atoms}
\end{figure*}

Here, we explore with a fully  ab-initio framework the Friedel oscillations emanating from single magnetic skyrmions hosted in Fe covered by thin films of Pd deposited on Ir(111) surface (see Figure~\ref{Fig_atoms}a). The latter material has been demonstrated to host sub-5 nm skyrmions~\cite{Romming2013,Romming2015}. The stabilization of the skyrmions is enabled by the Dzyaloshinskii-Moriya interaction ~\cite{Dzyalosinkii,Moriya} and  the Heisenberg exchange interaction. As a scattering state, we choose one of the surface/interface states characterizing this substrate, as identified in a previous work~\cite{Bouhassoune2019}. These states were also investigated in the context of the spin-mixing magnetoresistance (XMR), which enables the detection of non-collinear spin-textures such as magnetic skyrmions with all-electrical means~\cite{Crum2015,Hanneken2015,Fernandes2020}.  The XMR finds its origin in the noncollinearity of the magnetic texture enhanced with spin-orbit interaction, which impacts the electronic structure and thus the current flowing in a perpendicular fashion between the surface and a nonmagnetic electrode.

We investigate the scattering properties of skyrmions as function of their size and depth underneath the surface  once the spin-textures are buried below Pd films of various thicknesses. As a probing energy, we choose 2.15 eV, which is crossed by an interface state characterized by a wave vector 2 nm$^{-1}$,  and find that both the wavelength and phase shifts of the induced charge oscillations are not affected by the skyrmion's size in contrast to their amplitude, which can be enhanced owing to interference effects. The apparent width of the skyrmions as extracted from their electronic signature is found to be larger than their real size. The ripples act thus as a magnifier of the skyrmions. Remarkably, the Friedel oscillations can be large enough to remotely identify  skyrmions in an all-electrical fashion, even if hidden under the surface or laterally away from the detecting electrode.

\section{Materials and Methods}

Our first-principles simulations are based on the full-potential Korringa-Kohn-Rostoker Green function (KKR) method~\cite{Bauer2013,Papanikolaou2002} including self-consistently spin-orbit interaction as implemented within density functional theory (DFT). Exchange and correlation effects are treated in the local spin-density approximation as parametrized by Vosko, Wilk and Nusair~\cite{Vosko1980}. The ferromagnetic substrate is modeled with slabs containing 44 layers:  Pd$_{n{\mathrm{MLs}}}$/Fe/Ir(111), with $n=1,2,...,7$ being the number of Pd  monolayers (MLs). On one side of the slab, PdFe layers are deposited with  vacuum layers on either sides  similarly to was done in~\cite{Bouhassoune2019}. The atomic positions were taken from ~\cite{Dupe2014,Crum2015,Bouhassoune2019} assuming an fcc-stacking of the layers. After determining the  potential of the  two-dimensional-ferromagnetic slabs, the corresponding Green functions $G_0$ were harvested and skyrmions were embedded in the collinear substrate.  The Dyson equation schematically written as $G=G_0+G_0\Delta VG$ is solved in order to obtain the skyrmionic-defect Green function $G$. $\Delta V$ represents the modified atomic potential as compared to that of the unperturbed slab. This is performed until reaching self-consistency in the spirit of DFT.

The self-consistent calculations are performed using an angular momentum cutoff of $l$ = 3 for the scattering wave functions and a grid of $30 \times 30$ k-points mesh for the sampling of the two-dimensional Brillouin zone. The complex-energy contour needed for the energy integration is rectangular and consists of 40 grid points including seven Matsubara frequencies. For the calculation of the local density of states (LDOS), the Brillouin zone is sampled with a $200 \times 200$ k-points mesh. The surface-projected band structure is calculated along the following high symmetry directions $\overline\Gamma \overline K$, $\overline K \overline M$ and $\overline M \overline\Gamma$.  Furthermore, to have an adequate interpretation and analysis of our results we have considered a semi-infinite Ir (111) substrate using the decimation technique~\cite{Moliner1986,Szunyogh1994} to avoid any size effect that can blind our conclusions. 

We have investigated three magnetic defects {of different sizes as illustrated in \mbox{Figure \ref{Fig_atoms}b}} denoted Skyrmion 1, 2 and 3, where a number of Fe magnetic moments are rotated and allowed to get their direction and electronic structure updated in the self-consistent scheme. Note that Skyrmions 1 consists of a single Fe magnetic moment fliped with respect to the collinear surrounding. Skyrmions 2 and 3 contain 19 and 37 Fe atoms respectively besides their nearest neighboring Pd and Ir atoms. Experimentally, it is expected that by increasing the magnitude of the magnetic field would reduce the size of the skyrmion, leading to the sequence Skyrmion 3, 2 and ultimately 1. After converging the different sized skyrmionic profiles, various quantities were calculated, with a focus on the Friedel charge oscillations. The latter are computed in larger areas, about 100 nm$^2$ around the skyrmions at a height of 0.445 nm above the surface {(see the vacuum site shown in  \mbox{Figure \ref{Fig_atoms}a})}. Following the Tersoff-Hamann approximation~\cite{Tersoff1983}, the obtained spectra are related to the experimental differential conductance measurable by scanning tunneling microscopy.

\section{Results and Discussion}

\subsection{Interface State}

\begin{figure*}[ht!]
	\includegraphics*[angle=0,width=1.\linewidth]{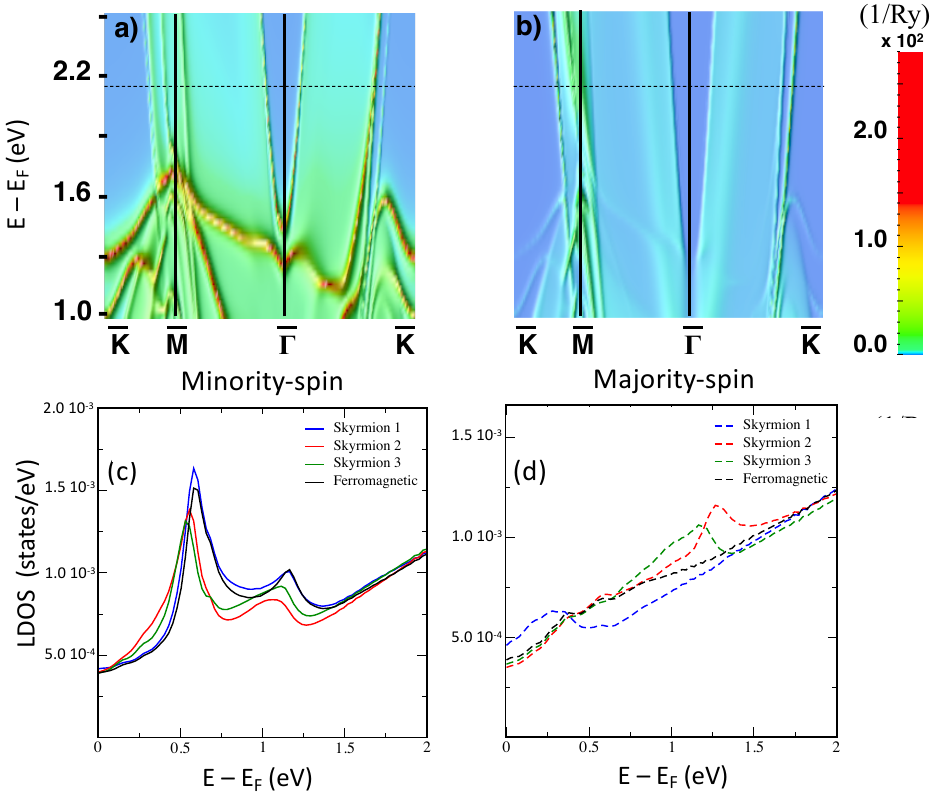}
	\caption{Electronic structure of PdFe bilayer deposited on Ir(111) surface with and without skyrmions. The layer-resolved surface-projected band structure of Fe is depicted for the minority- (\textbf{a}) and majority-spin (\textbf{b}) channels along high-symmetry directions of the two-dimensional Brillouin zone. The dashed line defines the energy, 2.15 eV, at which the Friedel oscillations are calculated. The Local density of states (LDOS) of the vacuum atop the core of the magnetic skyrmion for different skyrmion sizes compared to that of the ferromagnetic state for the minority- (\textbf{c}) and majority- (\textbf{d}) spin channels.}  

	\label{Fig2}
\end{figure*}

As shown in~\cite{Bouhassoune2019}, various interface and surface states and resonances are hosted by the PdFe bilayer deposited on Ir(111) surface. For our investigation, we consider the states living at 2.15 eV above the Fermi energy (see Figure~\ref{Fig2}). Here, an interface is located at the edge of the surface-projected bulk band gap besides other resonant states. The interface state is centered around the $\Gamma$ point in  reciprocal space  with a rather  quadratic dispersion (Figure~\ref{Fig2}a,b). At 2.15 eV it is characterized by a wave vector equal to 2 nm$^{-1}$. It is mainly of  minority-spin character and owing to spin-orbit coupling it is projected on the majority-spin channel~\cite{Bouhassoune2019}. 

The local density of states (LDOS) depicted in Figure~\ref{Fig2}c,d was obtained in the vacuum above  the PdFe bilayer  being either in a ferromagnetic state or hosting  confined spin-textures of different sizes: Skyrmion 1, 2 and 3 {illustrated in Figure~\ref{Fig_atoms}b}. {The magnetic texture is obviously not homogeneous and depends on the size of the skyrmions. For instance, the angle between the magnetic moment of the Fe atom at the skyrmion core and its nearest neighboring atoms decreases by increasing th size of the region confining the spin-texture. Beyond the nearest neighbors, the rotation angles do no change very much.} 
Although the spin is not a good quantum number, the LDOS is spin-resolved in the local atomic spin frame of reference at the vacuum site atop the core of the skyrmions.   One clearly sees that the vacuum LDOS hosts various resonances, already identified as surface or interface states~\cite{Bouhassoune2019},  distinctly showing up in either the majority- or minority-spin channels. Most important is that the LDOS is strongly dependent on the noncollinearity of the magnetic moments and on the size of the localized spin-texture. The interplay of the misalignement of the magnetic moments and hybridization of electronic states trigger the shift of the observed features and emergence of new states as discussed in~\cite{Crum2015,Bouhassoune2019}. This happens for instance in the minority-spin LDOS, which hosts new spectral feature in comparison to the ferromagnetic LDOS due to the hybridization and projection of the resonances occurring in the majority-spin channel.  

\subsection{Skyrmion-Induced Friedel Oscillations Induced in PdFe Bilayer on Ir(111)}

\begin{figure*}[ht!]
	\includegraphics*[angle=0,width=1.\linewidth]{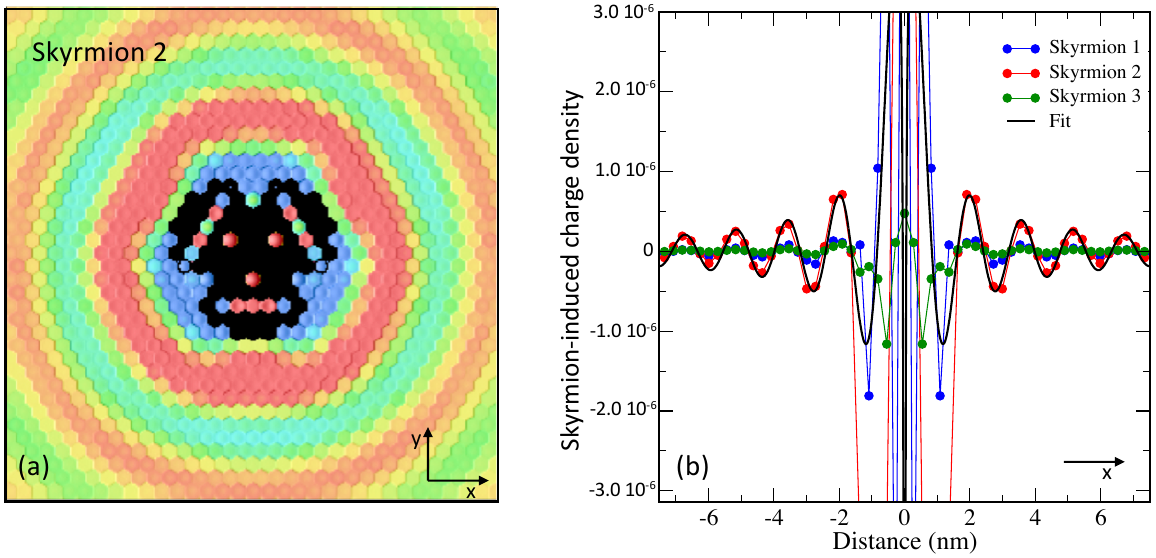}
	\caption{Skyrmion-induced charge oscillations in PdFe/Ir(111). (\textbf{a}) Ring-like ripples surrounding Skyrmion 2 at 2.15 eV. Atop the skyrmion the charge is extremely large and thus removed to observe the surrounding oscillations. 
	(\textbf{b}) Comparison of the Friedel charge waves along the x-direction as function of the size of the confined noncollinear spin-textures {shown in Figure~\ref{Fig_atoms}b}.}  
	\label{Fig3}
\end{figure*}

A magnetic skyrmion created in the PdFe/Ir(111) surface generates the Friedel oscillations shown in Figure~\ref{Fig3}a. The ring-like structures are not that isotropic due to the anisotropy of the constant energy surfaces characterizing the substrate's electrons participating in the scattering processes. A comparison of the charge ripples induced by the three investigated skyrmions at 2.15 eV along the x-axis is depicted in Figure~\ref{Fig3}b. Away from the skyrmions, the largest amplitude of the charge oscillations are induced by skyrmion 2. The wavelength of the ripples as well as their phase shifts seem to be the same for the three investigated noncollinear states. An important observation is that skyrmion 3, which is the largest in size, is leading to the most localized disturbance of the charge density. In other words, it would look like the smallest among the three localized spin-textures on the basis of Friedel oscillation. On the one hand, Skyrmion 1, which consists of a single flipped Fe spin-moment,  would appear more intense and wider than skyrmion 3. Skyrmion 2, on the other hand, would manifest as the largest skyrmion. This means that all-electrical detection enabled by scanning tunneling microscopy would distort the skyrmion and its real size, depending on how the charge perturbation propagates around the solitonic defect. The effect seen in the tunneling conductance is similar to the  distortion observed when looking at an object sitting under water. As discussed in the Appendix \ref{appendix}, various scattering mechanisms are at play when electrons scatter at complex spin-textures such as those investigated in the current work. Within the adiabatic approximation, i.e., assuming that the magnitude of the magnetic moment is not changed once rotated, it is expected in a first-order scattering approximation that the phase shift is independent from the noncollinearity of the magnetic moment. Interference effects intertwined with the magnetic inhomogeneity characterizing the environment of each atom within the confined spin-texture can either enhance or diminish the amplitude of the resulting charge oscillations. Remarkably, Skyrmion 2 seems to satisfy the right conditions to induce the largest magnitude of the Friedel ripples.

\subsection{Model for Phase Shifts and Mapping to a Simple-Scattering at an Atomic Defect}

The charge oscillations shown in Figure~\ref{Fig3} are similar to those induced by a point-defect inserted in a free electron gas. For the two-dimensional case, the modified change density in the asymptotic limit is expected to behave like:
\begin{eqnarray}
	\Delta n(r,\epsilon) &\propto& \frac{1}{r\sqrt{\epsilon}}
	\sin\left(
	2\sqrt{\epsilon} r+ \delta(\epsilon)
	\right) \, ,
\end{eqnarray}
in the s-wave approximation, which should be valid at large distances, $r$, away from the defect.  $\delta$ represents the phase shift experienced by the scattered wave functions. The fit to the induced charge oscillations plotted in Figure~\ref{Fig3}b is obtained with $\sqrt{\epsilon} = k = 2$ nm$^{-1}$, fitting perfectly the wave vector of the interface state located at 2.15 eV, and a phase shift of $\sim -\frac{\pi}{15}$, which is identical for the three investigated spin textures. We note that for the three investigated skyrmions, the wavelength of the scattered electronic state is larger than the size of the non-collinear region limited mainly to the Fe atom at the skyrmion's core and its nearest neighbors, which explains the obtained constant phase shift. Within scattering theory, a negative phase shift is the signature of an effective repulsive potential of the defect. The power decay of the oscillations confirm our assumption of the two-dimensionality of the scattering electrons. Naturally, the multiple scattering events occurring within the skyrmions are complex but at large distances from the solitonic defects, the scattering physics can be simplified and mapped to that known for  single atomic defects. The skyrmions can then be considered as particles with a characteristic scattering potential (repulsive in this case) and a phase shift. Similarly to atomic defects, the outcomes of a very simple scattering theory can be used to describe not only scattering aspects but also interaction patterns among solitonic defects, repulsive versus attraction,  between solitons and regular impurities, their spatial ordering. One could envisage very complex cases such those illustrated  schematically in Figure~\ref{Fig1}.

\subsection{All-Electrical Detection of Buried Skyrmions via Friedel Oscillations}

\begin{figure*}[ht!]
\includegraphics*[angle=0,width=1.\linewidth]{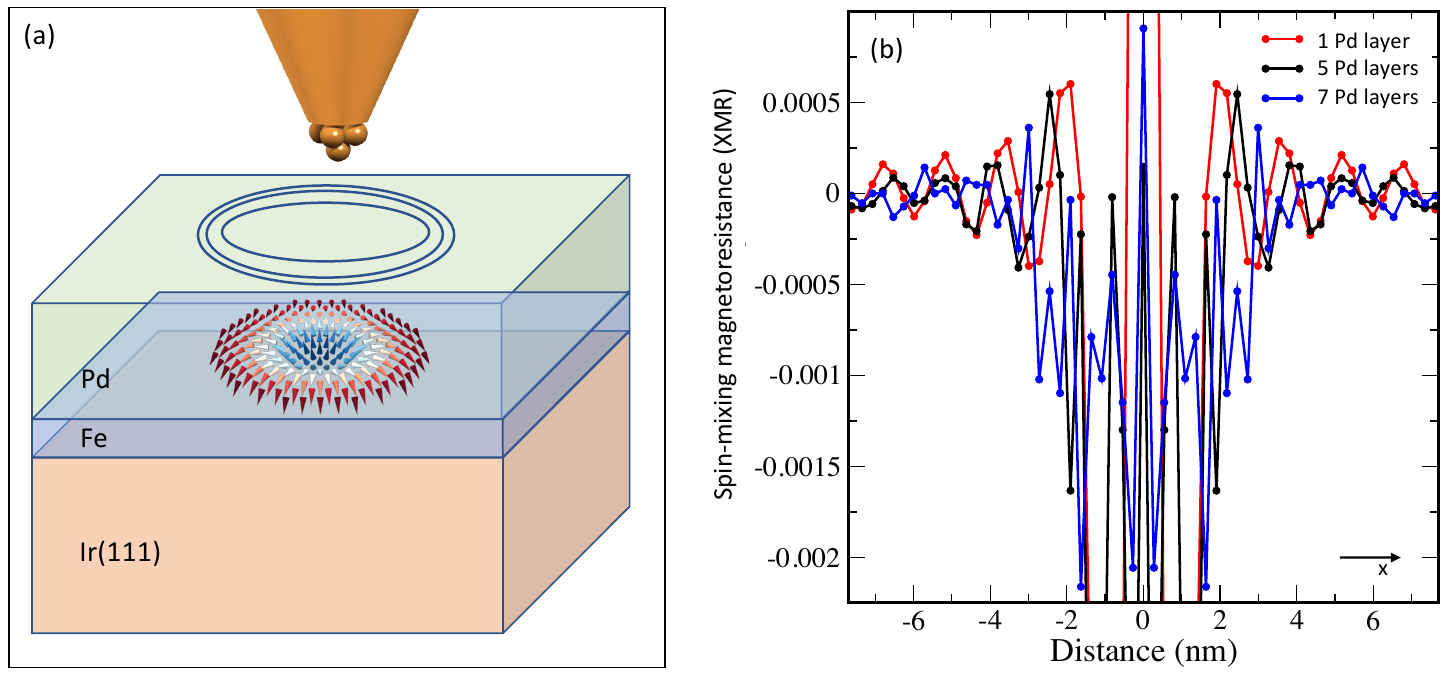}
	\caption{All\textnormal{-}electrical detection of skyrmions buried below a thin film of Pd. (\textbf{a}) Schematic picture representing the detection principle{, where a hidden skyrmion nanometers away from the surface can be sensed owing to the induced Friedel oscillations (blue circles) reaching the detecting electrode}; (\textbf{b}) spin-mixing magnetoresistance at 2.15 eV as function of the thickness of the Pd films. The lines are a guide for the eye.}  
	\label{Fig4}
\end{figure*}

As illustrated schematically in Figure~\ref{Fig4}a, we explore the case of skyrmions buried below an  overlayer film. In such a scenario, the detection of skyrmions is expected to be difficult. Here we demonstrate that Friedel oscillations can be useful in this context {since they can propagate and in some cases they can even be magnified through the spacer separating the skyrmions from the detecting electrode}. 
By dividing the induced charge oscillations by the charge density of the bare substrate, one obtains the Friedel XMR-signals plotted in Figure~\ref{Fig4}b along the x-direction for Skyrmion 2 buried under 1, 5 and 7 Pd layers. One clearly sees that the efficiency of the XMR-signal is of the same order of magnitude independently from the Pd film thickness. Moreover, the apparent size of the skyrmion seems to be the largest for the thickest Pd film. By increasing the thickness of the overlayer film, the positions of the surface and interface states characterizing the substrate can shift~\cite{Bouhassoune2019} besides  confinement effects coming into play accompanied with constructive or distructive interference mechanisms. Overall additional states are induced, which mediate and contribute  to the scattering of electrons at the solitonic defect. We note that on top of the Pd thick film, a surface state emerges, which helps to magnify the underlying Friedel oscillations emanating from the skyrmion located at the  PdFe interface.

 {The proposed detection mechanism is thus based on a nonmagnetic electrode that can scan the surface of the nonmagnetic overlayer film prospecting for the XMR-signal emanating from the Friedel oscillations induced by buried skyrmions. }
We conjecture that besides XMR other detection schemes~\cite{Palotas2021} can be used to detect the investigated Friedel~oscillations.

\section*{Conclusions}

To summarize, we performed ab-initio simulations on Friedel oscillations emanating from the scattering of electronic states at small magnetic skyrmions of different sizes. We explored the case of skyrmions hosted in an Fe layer deposited on Ir(111) surface all covered with Pd films with different thicknesses. We found that the wavelength and phase shift characterizing the induced charge ripples seem independent form the size of the skyrmions in strong contrast to their amplitude. This is intimately related to  interference effects of constructive and destructive nature in conjunction with the complexity of the spin-texture configurations. Through the Friedel oscillations, the skyrmion can appear larger or smaller than their real true size. The skyrmions act as particles to which a phase shift can be assigned, which permits the use of simple scattering theory  to explore scattering at, interaction and stability of complex configurations of skyrmions and other magnetic and atomic defects. The nature of the effective scattering potential characterizing the investigated skyrmions is found repulsive. We envisage that by tuning the skyrmion size, one can control the amplitude of the charge oscillations and therefore the interaction among skyrmions. This is not only important for the motion of skyrmions in racetrack devices but also in the detection and formation of lattices or superstructures made of confined spin-textures and defects.
Finally, we demonstrated  all-electrical detection of hidden skyrmions, or skyrmions buried below thick films, utilizing the XMR-signal characterizing the Friedel oscillations propagating through overlayers or spacers. Confinement effects can enhance the efficiency of the XMR signal and the overall appearance of the buried spin-textures.

 \section*{Acknowledgments}

This work was funded by the European Research Council (ERC) under the European Union's  Horizon 2020 research and innovation programme (ERC-consolidator Grant No. 681405 DYNASORE).
We gratefully acknowledge the computing time granted by the JARA-HPC Vergabegremium and VSR commission on the supercomputers JURECA at Forschungszentrum J\"ulich and CLAIX at RWTH-Aachen University.

\appendix*
\section{Multiple-scattering expansion for Friedel oscillations generated by a non-collinear magnetic state}

As mentioned in the introduction, magnetic skyrmions can be seen as defects living in a ferromagnetic collinear state. After scattering of electronic states at these defects, charge density oscillations are induced. Here, we derive the dependence of the induced charge density on the magnetic orientation of magnetic moments of the skyrmion. A formalism based on Green functions is very useful to describe the involved scattering processes. For simplicity, we neglect here the contribution of spin-orbit interaction.

We invoke a multiple scattering formalism~\cite{Papanikolaou2002} based on Green functions, $g$, describing the propagation of the electron states from site $i$ to site $j$. The theory is similar to the one utilized to explain the spin-mixing magnetoresistance in Ref.~\cite{Crum2015}. The  induced charge density at a site $i$  is given as
\begin{equation}
	\Delta n_i(\boldsymbol{r};\epsilon) = -\frac{1}{\pi}\mathrm{Im} 
	\,\trace
	[G_{ii}(\boldsymbol{r},\boldsymbol{r};\epsilon)-g_{ii}(\boldsymbol{r},\boldsymbol{r};\epsilon)]\, ,
	\label{LDOS}
\end{equation}
where the matrices ${g}$ and $G$ are the green functions of respectively the ferromagnetic host and of the skyrmion embedded in the host. The Green functions  can be represented in the angular momentum $L=(l,m)$ and spin representation, $L\oplus s$. Here, the trace Tr is taken over the 
spin indices. 

Using the Dyson equation, the change in the green function can be obtained from the series of Born iterations:
\begin{equation}
	G_{ii} = g_{ii} + \sum_j g_{ij} \Delta v_j g_{ji} + \sum_{jk} g_{ij} \Delta v_j g_{jk}\Delta v_k g_{ki} + ...\, ,
	\label{dyson}
\end{equation}
where $\Delta v$ represents the change of the atomic potential after rotating the corresponding magnetic moment. Note that for simplicity, the radial integrals accompanying the matrix products are not written explicitly.   

The potential pertaining to a given atom, $i$, can be written as:
\begin{equation}
	{v}_i (\boldsymbol{r})= \frac{1}{2}(v^{\uparrow}_i(\boldsymbol{r})+v^{\downarrow}_i(\boldsymbol{r}))\,\sigma_0+
	\frac{1}{2}(v^{\uparrow}_i(\boldsymbol{r})-v^{\downarrow}_i(\boldsymbol{r}))\, \mathbf{S}_i  \cdot \boldsymbol{\sigma},
\end{equation}
where $v^{\uparrow}$ and $v^{\downarrow}$ are the majority- and minority-spin $v$-matrices as obtained in the local-spin frame of reference of each atom. $\mathbf{S}_i$ is the unit vector of the spin magnetic moment at site $i$. Changing the orientation of the  magnetic moment of a given atom  and adopting the rigid spin approximation implies a corresponding change in the
spin-dependent $v$-matrix 
\begin{equation}
	\Delta {v}_i (\boldsymbol{r})= {v}^{\mathrm{s}}_i(\boldsymbol{r}) \, (\mathbf{S}_i -\mathbf{S}^z)  \cdot \boldsymbol{\sigma}\, ,
	\label{v_decomposed}
\end{equation}
in which ${v}^{\mathrm{s}}(\boldsymbol{r})=\frac{1}{2}(v^{\uparrow}_i(\boldsymbol{r})-v^{\downarrow}_i(\boldsymbol{r}))$ is the magnetic part of the total $v$-matrix.

The Green function of the ferromagnetic host that connects the atomic sites $i$ and $j$ can be partitioned into two  matrices:
\begin{equation}
	{g}_{ij} = {A}_{ij}\,\sigma_{0} 
	+ {{B}}_{ij}\,\mathbf{S}^z\cdot\boldsymbol{\sigma}\, ,
	\label{Green_decomposed}
\end{equation}
where ${A}_{ij}$ is the non-magnetic part of the Green function diagonal in spin space, while  ${{B}}_{ij}$ contains the magnetic part. $\boldsymbol{\sigma}$ is the vector of Pauli matrices and $\sigma_{0} $ is the identity matrix.  

We evaluate the induced charge density, Equation~(\ref{LDOS}), utilizing the \linebreak \mbox{Equations (\ref{dyson}), (\ref{v_decomposed}) and (\ref{Green_decomposed})} in combination with the useful properties of the Pauli matrices:
\begin{equation}
	\mathrm{Tr_S}\big[\sigma_x\big] = \mathrm{Tr_S}\big[\sigma_y\big] = \mathrm{Tr_S}\big[\sigma_z\big] =0 \;\; , 
\end{equation}
and
\begin{equation}
	(\mbox{\boldmath $\sigma$} \cdot \mathbf{\hat{S}})(\mbox{\boldmath $\sigma$} \cdot \mathbf{\hat{S}}^\prime) = \mathbf{\hat{S}} \cdot \mathbf{\hat{S}}^\prime + \mathrm{i} \ \mbox{\boldmath $\sigma$} \ \cdot  (\mathbf{\hat{S}} \times \mathbf{\hat{S}^\prime}) \; \;,
\end{equation}
where $\mathrm{i}$ is the imaginary unit. 

The first-order Born iteration yields:
\begin{equation}
	\Delta n^{(1)}_i = -\frac{4}{\pi}\mathrm{Im} 
	\,\trace
	\sum_j\, A_{ij}\, v_{j}^s\,B_{ji}\, (\mathbf{S}_j\cdot \mathbf{S}^z -1)
	\, ,
	\label{first-born}
\end{equation}
which leads to an angular dependence of the type $(1 -\cos{\theta_j})$, $\theta$ being the polar angle characterizing the moment. At this level, one can define an effective scattering matrix, $t(\epsilon) = t^s(\epsilon)  (\mathbf{S}_j\cdot \mathbf{S}^z -1)$ with $t^s(\epsilon) = \bra{\psi(\epsilon)} v^s \ket{\psi(\epsilon)}$, which describes how the magnetic part of the potential scatters the incoming electrons. The effective phase shift, $\delta(\epsilon)$,  that the Friedel oscillations would experience is related to the $t$-matrix by: $t(\epsilon) = \frac{1}{\sqrt{\epsilon}} \sin{\delta(\epsilon)} e^{\mathrm{i}\delta(\epsilon)}$. Therefore, $\tan{\delta(\epsilon)} = \frac{\Im t(\epsilon)}{\Re t(\epsilon)} = \frac{\Im t^s(\epsilon)}{\Re t^s(\epsilon)}$. In other words, within the firs-order Born iteration the amplitude of the induced Friedel oscillations, but not their phase shift,  would be affected by the rotation of the magnetic moment. This is resulting from the adiabatic approximation assuming the magnetic moment is not changed upon its rotation.

The second-order Born iteration generates more complex terms
\begin{widetext}
\begin{eqnarray}
\begin{array}{lll}
	\Delta n^{(2)}_i &=& -\frac{1}{\pi}\mathrm{Im} 
	\,\trace
	\sum_{jk}\,\big[ 
	(
	A_{ij}\,v_j^s\,A_{jk}\,v_k^s\,A_{ki} 
	+
	B_{ij}\,v_j^s\,B_{jk}\,v_k^s\,B_{ki}
	)
	(\boldsymbol{S}_j\cdot \boldsymbol{S}_k) \\ [1ex]
	&+&(
	B_{ij}\,v_j^s\,A_{jk}\,v_k^s\,B_{ki} 
	-B_{ij}\,v_j^s\,B_{jk}\,v_k^s\,A_{ki}
	-B_{ij}\,v_j^s\,B_{jk}\,v_k^s\,B_{ki}
	)(\boldsymbol{S}_j\cdot \boldsymbol{S}^z) (\boldsymbol{S}_k\cdot \boldsymbol{S}^z)\\ [1ex]
	&+&(
	A_{ij}\,v_j^s\,A_{jk}\,v_k^s\,A_{ki} 
	+B_{ij}\,v_j^s\,A_{jk}\,v_k^s\,B_{ki}
	-B_{ij}\,v_j^s\,B_{jk}\,v_k^s\,A_{ki}
	)(1- (\boldsymbol{S}_j+\boldsymbol{S}_k)\cdot \boldsymbol{S}^z) 
	\big]
	\, ,
	\label{second-born}
\end{array}\end{eqnarray}
\end{widetext}
yielding various angular dependencies, ranging from $\cos{\theta}$ to $\cos^2{\theta}$.

To summarize, the induced charge density is expected to have a complex angular dependence with respect to the orientation of the magnetic moments. There will be constructive and destructive interferences, which shape the amplitude of the charge ripples.

\bibliography{biblio}

\begin{thebibliography}{68}%
\makeatletter
\providecommand \@ifxundefined [1]{%
 \@ifx{#1\undefined}
}%
\providecommand \@ifnum [1]{%
 \ifnum #1\expandafter \@firstoftwo
 \else \expandafter \@secondoftwo
 \fi
}%
\providecommand \@ifx [1]{%
 \ifx #1\expandafter \@firstoftwo
 \else \expandafter \@secondoftwo
 \fi
}%
\providecommand \natexlab [1]{#1}%
\providecommand \enquote  [1]{``#1''}%
\providecommand \bibnamefont  [1]{#1}%
\providecommand \bibfnamefont [1]{#1}%
\providecommand \citenamefont [1]{#1}%
\providecommand \href@noop [0]{\@secondoftwo}%
\providecommand \href [0]{\begingroup \@sanitize@url \@href}%
\providecommand \@href[1]{\@@startlink{#1}\@@href}%
\providecommand \@@href[1]{\endgroup#1\@@endlink}%
\providecommand \@sanitize@url [0]{\catcode `\\12\catcode `\$12\catcode
  `\&12\catcode `\#12\catcode `\^12\catcode `\_12\catcode `\%12\relax}%
\providecommand \@@startlink[1]{}%
\providecommand \@@endlink[0]{}%
\providecommand \url  [0]{\begingroup\@sanitize@url \@url }%
\providecommand \@url [1]{\endgroup\@href {#1}{\urlprefix }}%
\providecommand \urlprefix  [0]{URL }%
\providecommand \Eprint [0]{\href }%
\providecommand \doibase [0]{http://dx.doi.org/}%
\providecommand \selectlanguage [0]{\@gobble}%
\providecommand \bibinfo  [0]{\@secondoftwo}%
\providecommand \bibfield  [0]{\@secondoftwo}%
\providecommand \translation [1]{[#1]}%
\providecommand \BibitemOpen [0]{}%
\providecommand \bibitemStop [0]{}%
\providecommand \bibitemNoStop [0]{.\EOS\space}%
\providecommand \EOS [0]{\spacefactor3000\relax}%
\providecommand \BibitemShut  [1]{\csname bibitem#1\endcsname}%
\let\auto@bib@innerbib\@empty
\bibitem [{\citenamefont {Crommie}\ \emph
  {et~al.}(1993{\natexlab{a}})\citenamefont {Crommie}, \citenamefont {Lutz},\
  and\ \citenamefont {Eigler}}]{Crommie1993}%
  \BibitemOpen
  \bibfield  {author} {\bibinfo {author} {\bibfnamefont {M.~F.}\ \bibnamefont
  {Crommie}}, \bibinfo {author} {\bibfnamefont {C.~P.}\ \bibnamefont {Lutz}}, \
  and\ \bibinfo {author} {\bibfnamefont {D.~M.}\ \bibnamefont {Eigler}},\
  }\bibfield  {title} {\enquote {\bibinfo {title} {Imaging standing waves in a
  two-dimensional electron gas},}\ }\href {\doibase 10.1038/363524a0}
  {\bibfield  {journal} {\bibinfo  {journal} {Nature}\ }\textbf {\bibinfo
  {volume} {363}},\ \bibinfo {pages} {524--527} (\bibinfo {year}
  {1993}{\natexlab{a}})}\BibitemShut {NoStop}%
\bibitem [{\citenamefont {Hasegawa}\ and\ \citenamefont
  {Avouris}(1993)}]{Hasegawa1993}%
  \BibitemOpen
  \bibfield  {author} {\bibinfo {author} {\bibfnamefont {Y.}~\bibnamefont
  {Hasegawa}}\ and\ \bibinfo {author} {\bibfnamefont {Ph.}\ \bibnamefont
  {Avouris}},\ }\bibfield  {title} {\enquote {\bibinfo {title} {Direct
  observation of standing wave formation at surface steps using scanning
  tunneling spectroscopy},}\ }\href {\doibase 10.1103/PhysRevLett.71.1071}
  {\bibfield  {journal} {\bibinfo  {journal} {Phys. Rev. Lett.}\ }\textbf
  {\bibinfo {volume} {71}},\ \bibinfo {pages} {1071--1074} (\bibinfo {year}
  {1993})}\BibitemShut {NoStop}%
\bibitem [{\citenamefont {Crommie}\ \emph
  {et~al.}(1993{\natexlab{b}})\citenamefont {Crommie}, \citenamefont {Lutz},\
  and\ \citenamefont {Eigler}}]{Crommie1993b}%
  \BibitemOpen
  \bibfield  {author} {\bibinfo {author} {\bibfnamefont {M.~F.}\ \bibnamefont
  {Crommie}}, \bibinfo {author} {\bibfnamefont {C.~P.}\ \bibnamefont {Lutz}}, \
  and\ \bibinfo {author} {\bibfnamefont {D.~M.}\ \bibnamefont {Eigler}},\
  }\bibfield  {title} {\enquote {\bibinfo {title} {Confinement of electrons to
  quantum corrals on a metal surface},}\ }\href {\doibase
  10.1126/science.262.5131.218} {\bibfield  {journal} {\bibinfo  {journal}
  {Science}\ }\textbf {\bibinfo {volume} {262}},\ \bibinfo {pages} {218--220}
  (\bibinfo {year} {1993}{\natexlab{b}})},\ \Eprint
  {http://arxiv.org/abs/https://science.sciencemag.org/content/262/5131/218.full.pdf}
  {https://science.sciencemag.org/content/262/5131/218.full.pdf} \BibitemShut
  {NoStop}%
\bibitem [{\citenamefont {Silly}\ \emph {et~al.}(2004)\citenamefont {Silly},
  \citenamefont {Pivetta}, \citenamefont {Ternes}, \citenamefont {Patthey},
  \citenamefont {Pelz},\ and\ \citenamefont {Schneider}}]{Silly2004}%
  \BibitemOpen
  \bibfield  {author} {\bibinfo {author} {\bibfnamefont {Fabien}\ \bibnamefont
  {Silly}}, \bibinfo {author} {\bibfnamefont {Marina}\ \bibnamefont {Pivetta}},
  \bibinfo {author} {\bibfnamefont {Markus}\ \bibnamefont {Ternes}}, \bibinfo
  {author} {\bibfnamefont {Fran\ifmmode \mbox{\c{c}}\else~\c{c}\fi{}ois}\
  \bibnamefont {Patthey}}, \bibinfo {author} {\bibfnamefont {Jonathan~P.}\
  \bibnamefont {Pelz}}, \ and\ \bibinfo {author} {\bibfnamefont {Wolf-Dieter}\
  \bibnamefont {Schneider}},\ }\bibfield  {title} {\enquote {\bibinfo {title}
  {Creation of an atomic superlattice by immersing metallic adatoms in a
  two-dimensional electron sea},}\ }\href {\doibase
  10.1103/PhysRevLett.92.016101} {\bibfield  {journal} {\bibinfo  {journal}
  {Phys. Rev. Lett.}\ }\textbf {\bibinfo {volume} {92}},\ \bibinfo {pages}
  {016101} (\bibinfo {year} {2004})}\BibitemShut {NoStop}%
\bibitem [{\citenamefont {Lounis}\ \emph {et~al.}(2012)\citenamefont {Lounis},
  \citenamefont {Bringer},\ and\ \citenamefont {Bl\"ugel}}]{Lounis:2012}%
  \BibitemOpen
  \bibfield  {author} {\bibinfo {author} {\bibfnamefont {Samir}\ \bibnamefont
  {Lounis}}, \bibinfo {author} {\bibfnamefont {Andreas}\ \bibnamefont
  {Bringer}}, \ and\ \bibinfo {author} {\bibfnamefont {Stefan}\ \bibnamefont
  {Bl\"ugel}},\ }\bibfield  {title} {\enquote {\bibinfo {title} {Magnetic
  adatom induced skyrmion-like spin texture in surface electron waves},}\
  }\href {\doibase 10.1103/PhysRevLett.108.207202} {\bibfield  {journal}
  {\bibinfo  {journal} {Phys. Rev. Lett.}\ }\textbf {\bibinfo {volume} {108}},\
  \bibinfo {pages} {207202} (\bibinfo {year} {2012})}\BibitemShut {NoStop}%
\bibitem [{\citenamefont {Meier}\ \emph {et~al.}(2011)\citenamefont {Meier},
  \citenamefont {Lounis}, \citenamefont {Wiebe}, \citenamefont {Zhou},
  \citenamefont {Heers}, \citenamefont {Mavropoulos}, \citenamefont
  {Dederichs}, \citenamefont {Bl\"ugel},\ and\ \citenamefont
  {Wiesendanger}}]{Meier2011}%
  \BibitemOpen
  \bibfield  {author} {\bibinfo {author} {\bibfnamefont {Focko}\ \bibnamefont
  {Meier}}, \bibinfo {author} {\bibfnamefont {Samir}\ \bibnamefont {Lounis}},
  \bibinfo {author} {\bibfnamefont {Jens}\ \bibnamefont {Wiebe}}, \bibinfo
  {author} {\bibfnamefont {Lihui}\ \bibnamefont {Zhou}}, \bibinfo {author}
  {\bibfnamefont {Swantje}\ \bibnamefont {Heers}}, \bibinfo {author}
  {\bibfnamefont {Phivos}\ \bibnamefont {Mavropoulos}}, \bibinfo {author}
  {\bibfnamefont {Peter~H.}\ \bibnamefont {Dederichs}}, \bibinfo {author}
  {\bibfnamefont {Stefan}\ \bibnamefont {Bl\"ugel}}, \ and\ \bibinfo {author}
  {\bibfnamefont {Roland}\ \bibnamefont {Wiesendanger}},\ }\bibfield  {title}
  {\enquote {\bibinfo {title} {Spin polarization of platinum (111) induced by
  the proximity to cobalt nanostripes},}\ }\href {\doibase
  10.1103/PhysRevB.83.075407} {\bibfield  {journal} {\bibinfo  {journal} {Phys.
  Rev. B}\ }\textbf {\bibinfo {volume} {83}},\ \bibinfo {pages} {075407}
  (\bibinfo {year} {2011})}\BibitemShut {NoStop}%
\bibitem [{\citenamefont {Weismann}\ \emph {et~al.}(2009)\citenamefont
  {Weismann}, \citenamefont {Wenderoth}, \citenamefont {Lounis}, \citenamefont
  {Zahn}, \citenamefont {Quaas}, \citenamefont {Ulbrich}, \citenamefont
  {Dederichs},\ and\ \citenamefont {Bl{\"u}gel}}]{Weismann2009}%
  \BibitemOpen
  \bibfield  {author} {\bibinfo {author} {\bibfnamefont {Alexander}\
  \bibnamefont {Weismann}}, \bibinfo {author} {\bibfnamefont {Martin}\
  \bibnamefont {Wenderoth}}, \bibinfo {author} {\bibfnamefont {Samir}\
  \bibnamefont {Lounis}}, \bibinfo {author} {\bibfnamefont {Peter}\
  \bibnamefont {Zahn}}, \bibinfo {author} {\bibfnamefont {Norbert}\
  \bibnamefont {Quaas}}, \bibinfo {author} {\bibfnamefont {Rainer~G.}\
  \bibnamefont {Ulbrich}}, \bibinfo {author} {\bibfnamefont {Peter~H.}\
  \bibnamefont {Dederichs}}, \ and\ \bibinfo {author} {\bibfnamefont {Stefan}\
  \bibnamefont {Bl{\"u}gel}},\ }\bibfield  {title} {\enquote {\bibinfo {title}
  {Seeing the fermi surface in real space by nanoscale electron focusing},}\
  }\href {\doibase 10.1126/science.1168738} {\bibfield  {journal} {\bibinfo
  {journal} {Science}\ }\textbf {\bibinfo {volume} {323}},\ \bibinfo {pages}
  {1190--1193} (\bibinfo {year} {2009})},\ \Eprint
  {http://arxiv.org/abs/https://science.sciencemag.org/content/323/5918/1190.full.pdf}
  {https://science.sciencemag.org/content/323/5918/1190.full.pdf} \BibitemShut
  {NoStop}%
\bibitem [{\citenamefont {Lounis}\ \emph {et~al.}(2011)\citenamefont {Lounis},
  \citenamefont {Zahn}, \citenamefont {Weismann}, \citenamefont {Wenderoth},
  \citenamefont {Ulbrich}, \citenamefont {Mertig}, \citenamefont {Dederichs},\
  and\ \citenamefont {Bl\"ugel}}]{Lounis2011}%
  \BibitemOpen
  \bibfield  {author} {\bibinfo {author} {\bibfnamefont {Samir}\ \bibnamefont
  {Lounis}}, \bibinfo {author} {\bibfnamefont {Peter}\ \bibnamefont {Zahn}},
  \bibinfo {author} {\bibfnamefont {Alexander}\ \bibnamefont {Weismann}},
  \bibinfo {author} {\bibfnamefont {Martin}\ \bibnamefont {Wenderoth}},
  \bibinfo {author} {\bibfnamefont {Rainer~G.}\ \bibnamefont {Ulbrich}},
  \bibinfo {author} {\bibfnamefont {Ingrid}\ \bibnamefont {Mertig}}, \bibinfo
  {author} {\bibfnamefont {Peter~H.}\ \bibnamefont {Dederichs}}, \ and\
  \bibinfo {author} {\bibfnamefont {Stefan}\ \bibnamefont {Bl\"ugel}},\
  }\bibfield  {title} {\enquote {\bibinfo {title} {Theory of real space imaging
  of fermi surface parts},}\ }\href {\doibase 10.1103/PhysRevB.83.035427}
  {\bibfield  {journal} {\bibinfo  {journal} {Phys. Rev. B}\ }\textbf {\bibinfo
  {volume} {83}},\ \bibinfo {pages} {035427} (\bibinfo {year}
  {2011})}\BibitemShut {NoStop}%
\bibitem [{\citenamefont {Pr{\"u}ser}\ \emph {et~al.}(2011)\citenamefont
  {Pr{\"u}ser}, \citenamefont {Wenderoth}, \citenamefont {Dargel},
  \citenamefont {Weismann}, \citenamefont {Peters}, \citenamefont {Pruschke},\
  and\ \citenamefont {Ulbrich}}]{Prueser2011}%
  \BibitemOpen
  \bibfield  {author} {\bibinfo {author} {\bibfnamefont {Henning}\ \bibnamefont
  {Pr{\"u}ser}}, \bibinfo {author} {\bibfnamefont {Martin}\ \bibnamefont
  {Wenderoth}}, \bibinfo {author} {\bibfnamefont {Piet~E.}\ \bibnamefont
  {Dargel}}, \bibinfo {author} {\bibfnamefont {Alexander}\ \bibnamefont
  {Weismann}}, \bibinfo {author} {\bibfnamefont {Robert}\ \bibnamefont
  {Peters}}, \bibinfo {author} {\bibfnamefont {Thomas}\ \bibnamefont
  {Pruschke}}, \ and\ \bibinfo {author} {\bibfnamefont {Rainer~G.}\
  \bibnamefont {Ulbrich}},\ }\bibfield  {title} {\enquote {\bibinfo {title}
  {Long-range kondo signature of a single magnetic impurity},}\ }\href
  {\doibase 10.1038/nphys1876} {\bibfield  {journal} {\bibinfo  {journal}
  {Nature Physics}\ }\textbf {\bibinfo {volume} {7}},\ \bibinfo {pages}
  {203--206} (\bibinfo {year} {2011})}\BibitemShut {NoStop}%
\bibitem [{\citenamefont {Pr\"user}\ \emph {et~al.}(2012)\citenamefont
  {Pr\"user}, \citenamefont {Wenderoth}, \citenamefont {Weismann},\ and\
  \citenamefont {Ulbrich}}]{Prueser2012}%
  \BibitemOpen
  \bibfield  {author} {\bibinfo {author} {\bibfnamefont {H.}~\bibnamefont
  {Pr\"user}}, \bibinfo {author} {\bibfnamefont {M.}~\bibnamefont {Wenderoth}},
  \bibinfo {author} {\bibfnamefont {A.}~\bibnamefont {Weismann}}, \ and\
  \bibinfo {author} {\bibfnamefont {R.~G.}\ \bibnamefont {Ulbrich}},\
  }\bibfield  {title} {\enquote {\bibinfo {title} {Mapping itinerant electrons
  around kondo impurities},}\ }\href {\doibase 10.1103/PhysRevLett.108.166604}
  {\bibfield  {journal} {\bibinfo  {journal} {Phys. Rev. Lett.}\ }\textbf
  {\bibinfo {volume} {108}},\ \bibinfo {pages} {166604} (\bibinfo {year}
  {2012})}\BibitemShut {NoStop}%
\bibitem [{\citenamefont {Avotina}\ \emph {et~al.}(2006)\citenamefont
  {Avotina}, \citenamefont {Kolesnichenko}, \citenamefont {Otte},\ and\
  \citenamefont {van Ruitenbeek}}]{Avotina2006}%
  \BibitemOpen
  \bibfield  {author} {\bibinfo {author} {\bibfnamefont {Ye.~S.}\ \bibnamefont
  {Avotina}}, \bibinfo {author} {\bibfnamefont {Yu.~A.}\ \bibnamefont
  {Kolesnichenko}}, \bibinfo {author} {\bibfnamefont {A.~F.}\ \bibnamefont
  {Otte}}, \ and\ \bibinfo {author} {\bibfnamefont {J.~M.}\ \bibnamefont {van
  Ruitenbeek}},\ }\bibfield  {title} {\enquote {\bibinfo {title} {Signature of
  fermi-surface anisotropy in point contact conductance in the presence of
  defects},}\ }\href {\doibase 10.1103/PhysRevB.74.085411} {\bibfield
  {journal} {\bibinfo  {journal} {Phys. Rev. B}\ }\textbf {\bibinfo {volume}
  {74}},\ \bibinfo {pages} {085411} (\bibinfo {year} {2006})}\BibitemShut
  {NoStop}%
\bibitem [{\citenamefont {Bouhassoune}\ \emph {et~al.}(2014)\citenamefont
  {Bouhassoune}, \citenamefont {Zimmermann}, \citenamefont {Mavropoulos},
  \citenamefont {Wortmann}, \citenamefont {Dederichs}, \citenamefont
  {Bl{\"u}gel},\ and\ \citenamefont {Lounis}}]{Bouhassoune2014}%
  \BibitemOpen
  \bibfield  {author} {\bibinfo {author} {\bibfnamefont {Mohammed}\
  \bibnamefont {Bouhassoune}}, \bibinfo {author} {\bibfnamefont {Bernd}\
  \bibnamefont {Zimmermann}}, \bibinfo {author} {\bibfnamefont {Phivos}\
  \bibnamefont {Mavropoulos}}, \bibinfo {author} {\bibfnamefont {Daniel}\
  \bibnamefont {Wortmann}}, \bibinfo {author} {\bibfnamefont {Peter~H.}\
  \bibnamefont {Dederichs}}, \bibinfo {author} {\bibfnamefont {Stefan}\
  \bibnamefont {Bl{\"u}gel}}, \ and\ \bibinfo {author} {\bibfnamefont {Samir}\
  \bibnamefont {Lounis}},\ }\bibfield  {title} {\enquote {\bibinfo {title}
  {Quantum well states and amplified spin-dependent friedel oscillations in
  thin films},}\ }\href {\doibase 10.1038/ncomms6558} {\bibfield  {journal}
  {\bibinfo  {journal} {Nature Communications}\ }\textbf {\bibinfo {volume}
  {5}},\ \bibinfo {pages} {5558} (\bibinfo {year} {2014})}\BibitemShut
  {NoStop}%
\bibitem [{\citenamefont {Kim}\ \emph {et~al.}(2020)\citenamefont {Kim},
  \citenamefont {R{\'o}zsa}, \citenamefont {Schreyer}, \citenamefont {Simon},\
  and\ \citenamefont {Wiesendanger}}]{Howon2020}%
  \BibitemOpen
  \bibfield  {author} {\bibinfo {author} {\bibfnamefont {Howon}\ \bibnamefont
  {Kim}}, \bibinfo {author} {\bibfnamefont {Levente}\ \bibnamefont
  {R{\'o}zsa}}, \bibinfo {author} {\bibfnamefont {Dominik}\ \bibnamefont
  {Schreyer}}, \bibinfo {author} {\bibfnamefont {Eszter}\ \bibnamefont
  {Simon}}, \ and\ \bibinfo {author} {\bibfnamefont {Roland}\ \bibnamefont
  {Wiesendanger}},\ }\bibfield  {title} {\enquote {\bibinfo {title} {Long-range
  focusing of magnetic bound states in superconducting lanthanum},}\ }\href
  {\doibase 10.1038/s41467-020-18406-8} {\bibfield  {journal} {\bibinfo
  {journal} {Nature Communications}\ }\textbf {\bibinfo {volume} {11}},\
  \bibinfo {pages} {4573} (\bibinfo {year} {2020})}\BibitemShut {NoStop}%
\bibitem [{\citenamefont {Zhou}\ \emph {et~al.}(2010)\citenamefont {Zhou},
  \citenamefont {Wiebe}, \citenamefont {Lounis}, \citenamefont {Vedmedenko},
  \citenamefont {Meier}, \citenamefont {Bl{\"u}gel}, \citenamefont
  {Dederichs},\ and\ \citenamefont {Wiesendanger}}]{Zhou2010}%
  \BibitemOpen
  \bibfield  {author} {\bibinfo {author} {\bibfnamefont {Lihui}\ \bibnamefont
  {Zhou}}, \bibinfo {author} {\bibfnamefont {Jens}\ \bibnamefont {Wiebe}},
  \bibinfo {author} {\bibfnamefont {Samir}\ \bibnamefont {Lounis}}, \bibinfo
  {author} {\bibfnamefont {Elena}\ \bibnamefont {Vedmedenko}}, \bibinfo
  {author} {\bibfnamefont {Focko}\ \bibnamefont {Meier}}, \bibinfo {author}
  {\bibfnamefont {Stefan}\ \bibnamefont {Bl{\"u}gel}}, \bibinfo {author}
  {\bibfnamefont {Peter~H.}\ \bibnamefont {Dederichs}}, \ and\ \bibinfo
  {author} {\bibfnamefont {Roland}\ \bibnamefont {Wiesendanger}},\ }\bibfield
  {title} {\enquote {\bibinfo {title} {Strength and directionality of surface
  ruderman--kittel--kasuya--yosida interaction mapped on the atomic scale},}\
  }\href {\doibase 10.1038/nphys1514} {\bibfield  {journal} {\bibinfo
  {journal} {Nature Physics}\ }\textbf {\bibinfo {volume} {6}},\ \bibinfo
  {pages} {187--191} (\bibinfo {year} {2010})}\BibitemShut {NoStop}%
\bibitem [{\citenamefont {Khajetoorians}\ \emph {et~al.}(2012)\citenamefont
  {Khajetoorians}, \citenamefont {Wiebe}, \citenamefont {Chilian},
  \citenamefont {Lounis}, \citenamefont {Bl{\"u}gel},\ and\ \citenamefont
  {Wiesendanger}}]{Khajetoorians2012}%
  \BibitemOpen
  \bibfield  {author} {\bibinfo {author} {\bibfnamefont {Alexander~Ako}\
  \bibnamefont {Khajetoorians}}, \bibinfo {author} {\bibfnamefont {Jens}\
  \bibnamefont {Wiebe}}, \bibinfo {author} {\bibfnamefont {Bruno}\ \bibnamefont
  {Chilian}}, \bibinfo {author} {\bibfnamefont {Samir}\ \bibnamefont {Lounis}},
  \bibinfo {author} {\bibfnamefont {Stefan}\ \bibnamefont {Bl{\"u}gel}}, \ and\
  \bibinfo {author} {\bibfnamefont {Roland}\ \bibnamefont {Wiesendanger}},\
  }\bibfield  {title} {\enquote {\bibinfo {title} {Atom-by-atom engineering and
  magnetometry of tailored nanomagnets},}\ }\href {\doibase 10.1038/nphys2299}
  {\bibfield  {journal} {\bibinfo  {journal} {Nature Physics}\ }\textbf
  {\bibinfo {volume} {8}},\ \bibinfo {pages} {497--503} (\bibinfo {year}
  {2012})}\BibitemShut {NoStop}%
\bibitem [{\citenamefont {Ngo}\ \emph {et~al.}(2012)\citenamefont {Ngo},
  \citenamefont {Rodriguez-Laguna}, \citenamefont {Ulloa},\ and\ \citenamefont
  {Kim}}]{Ngo2012}%
  \BibitemOpen
  \bibfield  {author} {\bibinfo {author} {\bibfnamefont {Anh~T.}\ \bibnamefont
  {Ngo}}, \bibinfo {author} {\bibfnamefont {Javier}\ \bibnamefont
  {Rodriguez-Laguna}}, \bibinfo {author} {\bibfnamefont {Sergio~E.}\
  \bibnamefont {Ulloa}}, \ and\ \bibinfo {author} {\bibfnamefont {Eugene~H.}\
  \bibnamefont {Kim}},\ }\bibfield  {title} {\enquote {\bibinfo {title}
  {Quantum manipulation via atomic-scale magnetoelectric effects},}\ }\bibfield
   {booktitle} {\emph {\bibinfo {booktitle} {Nano Letters}},\ }\href {\doibase
  10.1021/nl2025807} {\bibfield  {journal} {\bibinfo  {journal} {Nano Letters}\
  }\textbf {\bibinfo {volume} {12}},\ \bibinfo {pages} {13--16} (\bibinfo
  {year} {2012})}\BibitemShut {NoStop}%
\bibitem [{\citenamefont {Pr{\"u}ser}\ \emph {et~al.}(2014)\citenamefont
  {Pr{\"u}ser}, \citenamefont {Dargel}, \citenamefont {Bouhassoune},
  \citenamefont {Ulbrich}, \citenamefont {Pruschke}, \citenamefont {Lounis},\
  and\ \citenamefont {Wenderoth}}]{Prueser2014}%
  \BibitemOpen
  \bibfield  {author} {\bibinfo {author} {\bibfnamefont {Henning}\ \bibnamefont
  {Pr{\"u}ser}}, \bibinfo {author} {\bibfnamefont {Piet~E.}\ \bibnamefont
  {Dargel}}, \bibinfo {author} {\bibfnamefont {Mohammed}\ \bibnamefont
  {Bouhassoune}}, \bibinfo {author} {\bibfnamefont {Rainer~G.}\ \bibnamefont
  {Ulbrich}}, \bibinfo {author} {\bibfnamefont {Thomas}\ \bibnamefont
  {Pruschke}}, \bibinfo {author} {\bibfnamefont {Samir}\ \bibnamefont
  {Lounis}}, \ and\ \bibinfo {author} {\bibfnamefont {Martin}\ \bibnamefont
  {Wenderoth}},\ }\bibfield  {title} {\enquote {\bibinfo {title} {Interplay
  between the kondo effect and the ruderman--kittel--kasuya--yosida
  interaction},}\ }\href {\doibase 10.1038/ncomms6417} {\bibfield  {journal}
  {\bibinfo  {journal} {Nature Communications}\ }\textbf {\bibinfo {volume}
  {5}},\ \bibinfo {pages} {5417} (\bibinfo {year} {2014})}\BibitemShut
  {NoStop}%
\bibitem [{\citenamefont {Khajetoorians}\ \emph {et~al.}(2016)\citenamefont
  {Khajetoorians}, \citenamefont {Steinbrecher}, \citenamefont {Ternes},
  \citenamefont {Bouhassoune}, \citenamefont {dos Santos~Dias}, \citenamefont
  {Lounis}, \citenamefont {Wiebe},\ and\ \citenamefont
  {Wiesendanger}}]{Khajetoorians2016}%
  \BibitemOpen
  \bibfield  {author} {\bibinfo {author} {\bibfnamefont {A.~A.}\ \bibnamefont
  {Khajetoorians}}, \bibinfo {author} {\bibfnamefont {M.}~\bibnamefont
  {Steinbrecher}}, \bibinfo {author} {\bibfnamefont {M.}~\bibnamefont
  {Ternes}}, \bibinfo {author} {\bibfnamefont {M.}~\bibnamefont {Bouhassoune}},
  \bibinfo {author} {\bibfnamefont {M.}~\bibnamefont {dos Santos~Dias}},
  \bibinfo {author} {\bibfnamefont {S.}~\bibnamefont {Lounis}}, \bibinfo
  {author} {\bibfnamefont {J.}~\bibnamefont {Wiebe}}, \ and\ \bibinfo {author}
  {\bibfnamefont {R.}~\bibnamefont {Wiesendanger}},\ }\bibfield  {title}
  {\enquote {\bibinfo {title} {Tailoring the chiral magnetic interaction
  between two individual atoms},}\ }\href {\doibase 10.1038/ncomms10620}
  {\bibfield  {journal} {\bibinfo  {journal} {Nature Communications}\ }\textbf
  {\bibinfo {volume} {7}},\ \bibinfo {pages} {10620} (\bibinfo {year}
  {2016})}\BibitemShut {NoStop}%
\bibitem [{\citenamefont {Bouaziz}\ \emph {et~al.}(2020)\citenamefont
  {Bouaziz}, \citenamefont {Iba\~nez Azpiroz}, \citenamefont {Guimar\~aes},\
  and\ \citenamefont {Lounis}}]{Bouaziz2020}%
  \BibitemOpen
  \bibfield  {author} {\bibinfo {author} {\bibfnamefont {Juba}\ \bibnamefont
  {Bouaziz}}, \bibinfo {author} {\bibfnamefont {Julen}\ \bibnamefont {Iba\~nez
  Azpiroz}}, \bibinfo {author} {\bibfnamefont {Filipe S.~M.}\ \bibnamefont
  {Guimar\~aes}}, \ and\ \bibinfo {author} {\bibfnamefont {Samir}\ \bibnamefont
  {Lounis}},\ }\bibfield  {title} {\enquote {\bibinfo {title} {Zero-point
  magnetic exchange interactions},}\ }\href {\doibase
  10.1103/PhysRevResearch.2.043357} {\bibfield  {journal} {\bibinfo  {journal}
  {Phys. Rev. Research}\ }\textbf {\bibinfo {volume} {2}},\ \bibinfo {pages}
  {043357} (\bibinfo {year} {2020})}\BibitemShut {NoStop}%
\bibitem [{\citenamefont {Stepanyuk}\ \emph {et~al.}(2007)\citenamefont
  {Stepanyuk}, \citenamefont {Negulyaev}, \citenamefont {Niebergall},\ and\
  \citenamefont {Bruno}}]{Stepanyuk2007}%
  \BibitemOpen
  \bibfield  {author} {\bibinfo {author} {\bibfnamefont {V~S}\ \bibnamefont
  {Stepanyuk}}, \bibinfo {author} {\bibfnamefont {N~N}\ \bibnamefont
  {Negulyaev}}, \bibinfo {author} {\bibfnamefont {L}~\bibnamefont
  {Niebergall}}, \ and\ \bibinfo {author} {\bibfnamefont {P}~\bibnamefont
  {Bruno}},\ }\bibfield  {title} {\enquote {\bibinfo {title} {Effect of quantum
  confinement of surface electrons on adatom{\textendash}adatom
  interactions},}\ }\href {\doibase 10.1088/1367-2630/9/10/388} {\bibfield
  {journal} {\bibinfo  {journal} {New Journal of Physics}\ }\textbf {\bibinfo
  {volume} {9}},\ \bibinfo {pages} {388--388} (\bibinfo {year}
  {2007})}\BibitemShut {NoStop}%
\bibitem [{\citenamefont {Brovko}\ \emph {et~al.}(2008)\citenamefont {Brovko},
  \citenamefont {Ignatiev}, \citenamefont {Stepanyuk},\ and\ \citenamefont
  {Bruno}}]{Brovko2008}%
  \BibitemOpen
  \bibfield  {author} {\bibinfo {author} {\bibfnamefont {O.~O.}\ \bibnamefont
  {Brovko}}, \bibinfo {author} {\bibfnamefont {P.~A.}\ \bibnamefont
  {Ignatiev}}, \bibinfo {author} {\bibfnamefont {V.~S.}\ \bibnamefont
  {Stepanyuk}}, \ and\ \bibinfo {author} {\bibfnamefont {P.}~\bibnamefont
  {Bruno}},\ }\bibfield  {title} {\enquote {\bibinfo {title} {Tailoring
  exchange interactions in engineered nanostructures: An ab initio study},}\
  }\href {\doibase 10.1103/PhysRevLett.101.036809} {\bibfield  {journal}
  {\bibinfo  {journal} {Phys. Rev. Lett.}\ }\textbf {\bibinfo {volume} {101}},\
  \bibinfo {pages} {036809} (\bibinfo {year} {2008})}\BibitemShut {NoStop}%
\bibitem [{\citenamefont {Bogdanov}\ and\ \citenamefont
  {Yablonskii}(1989)}]{Bogdanov}%
  \BibitemOpen
  \bibfield  {author} {\bibinfo {author} {\bibfnamefont {Alexei~N}\
  \bibnamefont {Bogdanov}}\ and\ \bibinfo {author} {\bibfnamefont
  {DA}~\bibnamefont {Yablonskii}},\ }\bibfield  {title} {\enquote {\bibinfo
  {title} {Thermodynamically stable “vortices” in magnetically ordered
  crystals. the mixed state of magnets},}\ }\href@noop {} {\bibfield  {journal}
  {\bibinfo  {journal} {Zh. Eksp. Teor. Fiz}\ }\textbf {\bibinfo {volume}
  {95}},\ \bibinfo {pages} {178} (\bibinfo {year} {1989})}\BibitemShut
  {NoStop}%
\bibitem [{\citenamefont {R\"ossler}\ \emph {et~al.}(2006)\citenamefont
  {R\"ossler}, \citenamefont {Bogdanov},\ and\ \citenamefont
  {Pfleiderer}}]{Roessler2006}%
  \BibitemOpen
  \bibfield  {author} {\bibinfo {author} {\bibfnamefont {U.~K.}\ \bibnamefont
  {R\"ossler}}, \bibinfo {author} {\bibfnamefont {A.~N.}\ \bibnamefont
  {Bogdanov}}, \ and\ \bibinfo {author} {\bibfnamefont {C.}~\bibnamefont
  {Pfleiderer}},\ }\bibfield  {title} {\enquote {\bibinfo {title} {Spontaneous
  skyrmion ground states in magnetic metals},}\ }\href@noop {} {\bibfield
  {journal} {\bibinfo  {journal} {Nature}\ }\textbf {\bibinfo {volume} {442}},\
  \bibinfo {pages} {797--801} (\bibinfo {year} {2006})}\BibitemShut {NoStop}%
\bibitem [{\citenamefont {Nagaosa}\ and\ \citenamefont
  {Tokura}(2013)}]{Nagaosa2013}%
  \BibitemOpen
  \bibfield  {author} {\bibinfo {author} {\bibfnamefont {Naoto}\ \bibnamefont
  {Nagaosa}}\ and\ \bibinfo {author} {\bibfnamefont {Yoshinori}\ \bibnamefont
  {Tokura}},\ }\bibfield  {title} {\enquote {\bibinfo {title} {{Topological
  properties and dynamics of magnetic skyrmions.}}}\ }\href@noop {} {\bibfield
  {journal} {\bibinfo  {journal} {Nat. Nanotech.}\ }\textbf {\bibinfo {volume}
  {8}},\ \bibinfo {pages} {899--911} (\bibinfo {year} {2013})}\BibitemShut
  {NoStop}%
\bibitem [{\citenamefont {Fert}\ \emph {et~al.}(2013)\citenamefont {Fert},
  \citenamefont {Cros},\ and\ \citenamefont {Sampaio}}]{Fert2013}%
  \BibitemOpen
  \bibfield  {author} {\bibinfo {author} {\bibfnamefont {A.}~\bibnamefont
  {Fert}}, \bibinfo {author} {\bibfnamefont {V.}~\bibnamefont {Cros}}, \ and\
  \bibinfo {author} {\bibfnamefont {J.}~\bibnamefont {Sampaio}},\ }\bibfield
  {title} {\enquote {\bibinfo {title} {Skyrmions on the track.}}\ }\href@noop
  {} {\bibfield  {journal} {\bibinfo  {journal} {Nat Nano}\ ,\ \bibinfo {pages}
  {8(3):152–156}} (\bibinfo {year} {2013})}\BibitemShut {NoStop}%
\bibitem [{\citenamefont {Sampaio}\ \emph {et~al.}(2013)\citenamefont
  {Sampaio}, \citenamefont {Cros}, \citenamefont {Rohart}, \citenamefont
  {Thiaville},\ and\ \citenamefont {Fert}}]{Sampaio2013}%
  \BibitemOpen
  \bibfield  {author} {\bibinfo {author} {\bibfnamefont {J}~\bibnamefont
  {Sampaio}}, \bibinfo {author} {\bibfnamefont {V}~\bibnamefont {Cros}},
  \bibinfo {author} {\bibfnamefont {S}~\bibnamefont {Rohart}}, \bibinfo
  {author} {\bibfnamefont {A}~\bibnamefont {Thiaville}}, \ and\ \bibinfo
  {author} {\bibfnamefont {A}~\bibnamefont {Fert}},\ }\bibfield  {title}
  {\enquote {\bibinfo {title} {Nucleation, stability and current-induced motion
  of isolated magnetic skyrmions in nanostructures},}\ }\href@noop {}
  {\bibfield  {journal} {\bibinfo  {journal} {Nature Nanotech.}\ }\textbf
  {\bibinfo {volume} {8}},\ \bibinfo {pages} {839--844} (\bibinfo {year}
  {2013})}\BibitemShut {NoStop}%
\bibitem [{\citenamefont {Tomasello}\ \emph {et~al.}(2014)\citenamefont
  {Tomasello}, \citenamefont {Martinez}, \citenamefont {Zivieri}, \citenamefont
  {Torres}, \citenamefont {Carpentieri},\ and\ \citenamefont
  {Finocchio}}]{Tomasello2013}%
  \BibitemOpen
  \bibfield  {author} {\bibinfo {author} {\bibfnamefont {R}~\bibnamefont
  {Tomasello}}, \bibinfo {author} {\bibfnamefont {E}~\bibnamefont {Martinez}},
  \bibinfo {author} {\bibfnamefont {R}~\bibnamefont {Zivieri}}, \bibinfo
  {author} {\bibfnamefont {L}~\bibnamefont {Torres}}, \bibinfo {author}
  {\bibfnamefont {M}~\bibnamefont {Carpentieri}}, \ and\ \bibinfo {author}
  {\bibfnamefont {G}~\bibnamefont {Finocchio}},\ }\bibfield  {title} {\enquote
  {\bibinfo {title} {A strategy for the design of skyrmion racetrack
  memories},}\ }\href@noop {} {\bibfield  {journal} {\bibinfo  {journal}
  {Scientific Reports}\ }\textbf {\bibinfo {volume} {4}},\ \bibinfo {pages}
  {6784} (\bibinfo {year} {2014})}\BibitemShut {NoStop}%
\bibitem [{\citenamefont {Zhou}\ and\ \citenamefont {Ezawa}(2014)}]{Zhou2014}%
  \BibitemOpen
  \bibfield  {author} {\bibinfo {author} {\bibfnamefont {Yan}\ \bibnamefont
  {Zhou}}\ and\ \bibinfo {author} {\bibfnamefont {Motohiko}\ \bibnamefont
  {Ezawa}},\ }\bibfield  {title} {\enquote {\bibinfo {title} {{A reversible
  conversion between a skyrmion and a domain-wall pair in a junction
  geometry}},}\ }\href@noop {} {\bibfield  {journal} {\bibinfo  {journal}
  {Nature Commun.}\ }\textbf {\bibinfo {volume} {5}},\ \bibinfo {pages} {4652}
  (\bibinfo {year} {2014})}\BibitemShut {NoStop}%
\bibitem [{\citenamefont {Crum}\ \emph {et~al.}(2015)\citenamefont {Crum},
  \citenamefont {Bouhassoune}, \citenamefont {Bouaziz}, \citenamefont
  {Schweflinghaus}, \citenamefont {Bl{\"{u}}gel},\ and\ \citenamefont
  {Lounis}}]{Crum2015}%
  \BibitemOpen
  \bibfield  {author} {\bibinfo {author} {\bibfnamefont {Dax~M.}\ \bibnamefont
  {Crum}}, \bibinfo {author} {\bibfnamefont {Mohammed}\ \bibnamefont
  {Bouhassoune}}, \bibinfo {author} {\bibfnamefont {Juba}\ \bibnamefont
  {Bouaziz}}, \bibinfo {author} {\bibfnamefont {Benedikt}\ \bibnamefont
  {Schweflinghaus}}, \bibinfo {author} {\bibfnamefont {Stefan}\ \bibnamefont
  {Bl{\"{u}}gel}}, \ and\ \bibinfo {author} {\bibfnamefont {Samir}\
  \bibnamefont {Lounis}},\ }\bibfield  {title} {\enquote {\bibinfo {title}
  {{Perpendicular reading of single confined magnetic skyrmions}},}\
  }\href@noop {} {\bibfield  {journal} {\bibinfo  {journal} {Nature Commun.}\
  }\textbf {\bibinfo {volume} {6}},\ \bibinfo {pages} {8541} (\bibinfo {year}
  {2015})}\BibitemShut {NoStop}%
\bibitem [{\citenamefont {Zhang}\ \emph
  {et~al.}(2015{\natexlab{a}})\citenamefont {Zhang}, \citenamefont {Ezawa},\
  and\ \citenamefont {Zhou}}]{Zhang2015}%
  \BibitemOpen
  \bibfield  {author} {\bibinfo {author} {\bibfnamefont {Xichao}\ \bibnamefont
  {Zhang}}, \bibinfo {author} {\bibfnamefont {Motohiko}\ \bibnamefont {Ezawa}},
  \ and\ \bibinfo {author} {\bibfnamefont {Yan}\ \bibnamefont {Zhou}},\
  }\bibfield  {title} {\enquote {\bibinfo {title} {Magnetic skyrmion logic
  gates: conversion, duplication and merging of skyrmions},}\ }\href@noop {}
  {\bibfield  {journal} {\bibinfo  {journal} {Scientific Reports}\ }\textbf
  {\bibinfo {volume} {5}},\ \bibinfo {pages} {9400} (\bibinfo {year}
  {2015}{\natexlab{a}})}\BibitemShut {NoStop}%
\bibitem [{\citenamefont {Yu}\ \emph {et~al.}(2016)\citenamefont {Yu},
  \citenamefont {Upadhyaya}, \citenamefont {Li}, \citenamefont {Li},
  \citenamefont {Kim}, \citenamefont {Fan}, \citenamefont {Wong}, \citenamefont
  {Tserkovnyak}, \citenamefont {Amiri},\ and\ \citenamefont {Wang}}]{Yu2016}%
  \BibitemOpen
  \bibfield  {author} {\bibinfo {author} {\bibfnamefont {Guoqiang}\
  \bibnamefont {Yu}}, \bibinfo {author} {\bibfnamefont {Pramey}\ \bibnamefont
  {Upadhyaya}}, \bibinfo {author} {\bibfnamefont {Xiang}\ \bibnamefont {Li}},
  \bibinfo {author} {\bibfnamefont {Wenyuan}\ \bibnamefont {Li}}, \bibinfo
  {author} {\bibfnamefont {Se~Kwon}\ \bibnamefont {Kim}}, \bibinfo {author}
  {\bibfnamefont {Yabin}\ \bibnamefont {Fan}}, \bibinfo {author} {\bibfnamefont
  {Kin~L.}\ \bibnamefont {Wong}}, \bibinfo {author} {\bibfnamefont {Yaroslav}\
  \bibnamefont {Tserkovnyak}}, \bibinfo {author} {\bibfnamefont
  {Pedram~Khalili}\ \bibnamefont {Amiri}}, \ and\ \bibinfo {author}
  {\bibfnamefont {Kang~L.}\ \bibnamefont {Wang}},\ }\bibfield  {title}
  {\enquote {\bibinfo {title} {Room-temperature creation and spin-orbit torque
  manipulation of skyrmions in thin films with engineered asymmetry},}\
  }\href@noop {} {\bibfield  {journal} {\bibinfo  {journal} {Nano Letters}\
  }\textbf {\bibinfo {volume} {16}},\ \bibinfo {pages} {1981--1988} (\bibinfo
  {year} {2016})}\BibitemShut {NoStop}%
\bibitem [{\citenamefont {Garcia-Sanchez}\ \emph {et~al.}(2016)\citenamefont
  {Garcia-Sanchez}, \citenamefont {Sampaio}, \citenamefont {Reyren},
  \citenamefont {Cros},\ and\ \citenamefont {Kim}}]{Garcia-Sanchez2016}%
  \BibitemOpen
  \bibfield  {author} {\bibinfo {author} {\bibfnamefont {F}~\bibnamefont
  {Garcia-Sanchez}}, \bibinfo {author} {\bibfnamefont {J}~\bibnamefont
  {Sampaio}}, \bibinfo {author} {\bibfnamefont {N}~\bibnamefont {Reyren}},
  \bibinfo {author} {\bibfnamefont {V}~\bibnamefont {Cros}}, \ and\ \bibinfo
  {author} {\bibfnamefont {JV}~\bibnamefont {Kim}},\ }\bibfield  {title}
  {\enquote {\bibinfo {title} {A skyrmion-based spin-torque nano-oscillator},}\
  }\href@noop {} {\bibfield  {journal} {\bibinfo  {journal} {New Journal of
  Physics}\ }\textbf {\bibinfo {volume} {18}},\ \bibinfo {pages} {075011}
  (\bibinfo {year} {2016})}\BibitemShut {NoStop}%
\bibitem [{\citenamefont {Xia}\ \emph {et~al.}(2017)\citenamefont {Xia},
  \citenamefont {Jin}, \citenamefont {Song}, \citenamefont {Wang},
  \citenamefont {Wang},\ and\ \citenamefont {Liu}}]{Xia2017}%
  \BibitemOpen
  \bibfield  {author} {\bibinfo {author} {\bibfnamefont {Haiyan}\ \bibnamefont
  {Xia}}, \bibinfo {author} {\bibfnamefont {Chendong}\ \bibnamefont {Jin}},
  \bibinfo {author} {\bibfnamefont {Chengkun}\ \bibnamefont {Song}}, \bibinfo
  {author} {\bibfnamefont {Jinshuai}\ \bibnamefont {Wang}}, \bibinfo {author}
  {\bibfnamefont {Jianbo}\ \bibnamefont {Wang}}, \ and\ \bibinfo {author}
  {\bibfnamefont {Qingfang}\ \bibnamefont {Liu}},\ }\bibfield  {title}
  {\enquote {\bibinfo {title} {Control and manipulation of antiferromagnetic
  skyrmions in racetrack},}\ }\href {\doibase 10.1088/1361-6463/aa95f2}
  {\bibfield  {journal} {\bibinfo  {journal} {Journal of Physics D: Applied
  Physics}\ }\textbf {\bibinfo {volume} {50}},\ \bibinfo {pages} {505005}
  (\bibinfo {year} {2017})}\BibitemShut {NoStop}%
\bibitem [{\citenamefont {Fernandes}\ \emph
  {et~al.}(2020{\natexlab{a}})\citenamefont {Fernandes}, \citenamefont
  {Bouhassoune},\ and\ \citenamefont {Lounis}}]{Fernandes2020}%
  \BibitemOpen
  \bibfield  {author} {\bibinfo {author} {\bibfnamefont {Imara~Lima}\
  \bibnamefont {Fernandes}}, \bibinfo {author} {\bibfnamefont {Mohammed}\
  \bibnamefont {Bouhassoune}}, \ and\ \bibinfo {author} {\bibfnamefont {Samir}\
  \bibnamefont {Lounis}},\ }\bibfield  {title} {\enquote {\bibinfo {title}
  {Defect-implantation for the all-electrical detection of non-collinear
  spin-textures},}\ }\href@noop {} {\bibfield  {journal} {\bibinfo  {journal}
  {Nature Commun.}\ }\textbf {\bibinfo {volume} {11}},\ \bibinfo {pages} {1--9}
  (\bibinfo {year} {2020}{\natexlab{a}})}\BibitemShut {NoStop}%
\bibitem [{\citenamefont {Li}\ \emph {et~al.}(2021)\citenamefont {Li},
  \citenamefont {Kang}, \citenamefont {Zhang}, \citenamefont {Nie},
  \citenamefont {Zhou}, \citenamefont {Wang},\ and\ \citenamefont
  {Zhao}}]{Sai2021}%
  \BibitemOpen
  \bibfield  {author} {\bibinfo {author} {\bibfnamefont {Sai}\ \bibnamefont
  {Li}}, \bibinfo {author} {\bibfnamefont {Wang}\ \bibnamefont {Kang}},
  \bibinfo {author} {\bibfnamefont {Xichao}\ \bibnamefont {Zhang}}, \bibinfo
  {author} {\bibfnamefont {Tianxiao}\ \bibnamefont {Nie}}, \bibinfo {author}
  {\bibfnamefont {Yan}\ \bibnamefont {Zhou}}, \bibinfo {author} {\bibfnamefont
  {Kang~L.}\ \bibnamefont {Wang}}, \ and\ \bibinfo {author} {\bibfnamefont
  {Weisheng}\ \bibnamefont {Zhao}},\ }\bibfield  {title} {\enquote {\bibinfo
  {title} {Magnetic skyrmions for unconventional computing},}\ }\href {\doibase
  10.1039/D0MH01603A} {\bibfield  {journal} {\bibinfo  {journal} {Mater.
  Horiz.}\ ,\ \bibinfo {pages} {--}} (\bibinfo {year} {2021})}\BibitemShut
  {NoStop}%
\bibitem [{\citenamefont {Parkin}\ \emph {et~al.}(2008)\citenamefont {Parkin},
  \citenamefont {Hayashi},\ and\ \citenamefont {Thomas}}]{Parkin2008}%
  \BibitemOpen
  \bibfield  {author} {\bibinfo {author} {\bibfnamefont {Stuart S.~P.}\
  \bibnamefont {Parkin}}, \bibinfo {author} {\bibfnamefont {Masamitsu}\
  \bibnamefont {Hayashi}}, \ and\ \bibinfo {author} {\bibfnamefont {Luc}\
  \bibnamefont {Thomas}},\ }\bibfield  {title} {\enquote {\bibinfo {title}
  {Magnetic domain-wall racetrack memory},}\ }\href {\doibase
  10.1126/science.1145799} {\bibfield  {journal} {\bibinfo  {journal}
  {Science}\ }\textbf {\bibinfo {volume} {320}},\ \bibinfo {pages} {190--194}
  (\bibinfo {year} {2008})},\ \Eprint
  {http://arxiv.org/abs/https://science.sciencemag.org/content/320/5873/190.full.pdf}
  {https://science.sciencemag.org/content/320/5873/190.full.pdf} \BibitemShut
  {NoStop}%
\bibitem [{\citenamefont {Du}\ \emph {et~al.}(2018)\citenamefont {Du},
  \citenamefont {Zhao}, \citenamefont {Rybakov}, \citenamefont {Borisov},
  \citenamefont {Wang}, \citenamefont {Tang}, \citenamefont {Jin},
  \citenamefont {Wang}, \citenamefont {Wei}, \citenamefont {Kiselev},
  \citenamefont {Zhang}, \citenamefont {Che}, \citenamefont {Bl\"ugel},\ and\
  \citenamefont {Tian}}]{Haifeng2018}%
  \BibitemOpen
  \bibfield  {author} {\bibinfo {author} {\bibfnamefont {Haifeng}\ \bibnamefont
  {Du}}, \bibinfo {author} {\bibfnamefont {Xuebing}\ \bibnamefont {Zhao}},
  \bibinfo {author} {\bibfnamefont {Filipp~N.}\ \bibnamefont {Rybakov}},
  \bibinfo {author} {\bibfnamefont {Aleksandr~B.}\ \bibnamefont {Borisov}},
  \bibinfo {author} {\bibfnamefont {Shasha}\ \bibnamefont {Wang}}, \bibinfo
  {author} {\bibfnamefont {Jin}\ \bibnamefont {Tang}}, \bibinfo {author}
  {\bibfnamefont {Chiming}\ \bibnamefont {Jin}}, \bibinfo {author}
  {\bibfnamefont {Chao}\ \bibnamefont {Wang}}, \bibinfo {author} {\bibfnamefont
  {Wensheng}\ \bibnamefont {Wei}}, \bibinfo {author} {\bibfnamefont
  {Nikolai~S.}\ \bibnamefont {Kiselev}}, \bibinfo {author} {\bibfnamefont
  {Yuheng}\ \bibnamefont {Zhang}}, \bibinfo {author} {\bibfnamefont {Renchao}\
  \bibnamefont {Che}}, \bibinfo {author} {\bibfnamefont {Stefan}\ \bibnamefont
  {Bl\"ugel}}, \ and\ \bibinfo {author} {\bibfnamefont {Mingliang}\
  \bibnamefont {Tian}},\ }\bibfield  {title} {\enquote {\bibinfo {title}
  {Interaction of individual skyrmions in a nanostructured cubic chiral
  magnet},}\ }\href {\doibase 10.1103/PhysRevLett.120.197203} {\bibfield
  {journal} {\bibinfo  {journal} {Phys. Rev. Lett.}\ }\textbf {\bibinfo
  {volume} {120}},\ \bibinfo {pages} {197203} (\bibinfo {year}
  {2018})}\BibitemShut {NoStop}%
\bibitem [{\citenamefont {R\'ozsa}\ \emph {et~al.}(2016)\citenamefont
  {R\'ozsa}, \citenamefont {De\'ak}, \citenamefont {Simon}, \citenamefont
  {Yanes}, \citenamefont {Udvardi}, \citenamefont {Szunyogh},\ and\
  \citenamefont {Nowak}}]{Rozsa:2016}%
  \BibitemOpen
  \bibfield  {author} {\bibinfo {author} {\bibfnamefont {Levente}\ \bibnamefont
  {R\'ozsa}}, \bibinfo {author} {\bibfnamefont {Andr\'as}\ \bibnamefont
  {De\'ak}}, \bibinfo {author} {\bibfnamefont {Eszter}\ \bibnamefont {Simon}},
  \bibinfo {author} {\bibfnamefont {Rocio}\ \bibnamefont {Yanes}}, \bibinfo
  {author} {\bibfnamefont {L\'aszl\'o}\ \bibnamefont {Udvardi}}, \bibinfo
  {author} {\bibfnamefont {L\'aszl\'o}\ \bibnamefont {Szunyogh}}, \ and\
  \bibinfo {author} {\bibfnamefont {Ulrich}\ \bibnamefont {Nowak}},\ }\bibfield
   {title} {\enquote {\bibinfo {title} {Skyrmions with attractive interactions
  in an ultrathin magnetic film},}\ }\href {\doibase
  10.1103/PhysRevLett.117.157205} {\bibfield  {journal} {\bibinfo  {journal}
  {Phys. Rev. Lett.}\ }\textbf {\bibinfo {volume} {117}},\ \bibinfo {pages}
  {157205} (\bibinfo {year} {2016})}\BibitemShut {NoStop}%
\bibitem [{\citenamefont {Fernandes}\ \emph
  {et~al.}(2020{\natexlab{b}})\citenamefont {Fernandes}, \citenamefont
  {Chico},\ and\ \citenamefont {Lounis}}]{Fernandes2020a}%
  \BibitemOpen
  \bibfield  {author} {\bibinfo {author} {\bibfnamefont {Imara~Lima}\
  \bibnamefont {Fernandes}}, \bibinfo {author} {\bibfnamefont {Jonathan}\
  \bibnamefont {Chico}}, \ and\ \bibinfo {author} {\bibfnamefont {Samir}\
  \bibnamefont {Lounis}},\ }\bibfield  {title} {\enquote {\bibinfo {title}
  {{Impurity-dependent gyrotropic motion, deflection and pinning of
  current-driven ultrasmall skyrmions in PdFe/Ir(111) surface}},}\ }\href
  {http://iopscience.iop.org/10.1088/1361-648X/ab9cf0} {\bibfield  {journal}
  {\bibinfo  {journal} {Journal of Physics: Condensed Matter}\ } (\bibinfo
  {year} {2020}{\natexlab{b}})}\BibitemShut {NoStop}%
\bibitem [{\citenamefont {Arjana}\ \emph {et~al.}(2020)\citenamefont {Arjana},
  \citenamefont {Lima~Fernandes}, \citenamefont {Chico},\ and\ \citenamefont
  {Lounis}}]{Arjana2020}%
  \BibitemOpen
  \bibfield  {author} {\bibinfo {author} {\bibfnamefont {I.~Gede}\ \bibnamefont
  {Arjana}}, \bibinfo {author} {\bibfnamefont {Imara}\ \bibnamefont
  {Lima~Fernandes}}, \bibinfo {author} {\bibfnamefont {Jonathan}\ \bibnamefont
  {Chico}}, \ and\ \bibinfo {author} {\bibfnamefont {Samir}\ \bibnamefont
  {Lounis}},\ }\bibfield  {title} {\enquote {\bibinfo {title} {Sub-nanoscale
  atom-by-atom crafting of skyrmion-defect interaction profiles},}\ }\href
  {\doibase 10.1038/s41598-020-71232-2} {\bibfield  {journal} {\bibinfo
  {journal} {Scientific Reports}\ }\textbf {\bibinfo {volume} {10}},\ \bibinfo
  {pages} {14655} (\bibinfo {year} {2020})}\BibitemShut {NoStop}%
\bibitem [{\citenamefont {Huang}\ \emph {et~al.}(2017)\citenamefont {Huang},
  \citenamefont {Kang}, \citenamefont {Zhang}, \citenamefont {Zhou},\ and\
  \citenamefont {Zhao}}]{Huang2017}%
  \BibitemOpen
  \bibfield  {author} {\bibinfo {author} {\bibfnamefont {Yangqi}\ \bibnamefont
  {Huang}}, \bibinfo {author} {\bibfnamefont {Wang}\ \bibnamefont {Kang}},
  \bibinfo {author} {\bibfnamefont {Xichao}\ \bibnamefont {Zhang}}, \bibinfo
  {author} {\bibfnamefont {Yan}\ \bibnamefont {Zhou}}, \ and\ \bibinfo {author}
  {\bibfnamefont {Weisheng}\ \bibnamefont {Zhao}},\ }\bibfield  {title}
  {\enquote {\bibinfo {title} {Magnetic skyrmion-based synaptic devices},}\
  }\href {\doibase 10.1088/1361-6528/aa5838} {\bibfield  {journal} {\bibinfo
  {journal} {Nanotechnology}\ }\textbf {\bibinfo {volume} {28}},\ \bibinfo
  {pages} {08LT02} (\bibinfo {year} {2017})}\BibitemShut {NoStop}%
\bibitem [{\citenamefont {Song}\ \emph {et~al.}(2020)\citenamefont {Song},
  \citenamefont {Jeong}, \citenamefont {Pan}, \citenamefont {Zhang},
  \citenamefont {Xia}, \citenamefont {Cha}, \citenamefont {Park}, \citenamefont
  {Kim}, \citenamefont {Finizio}, \citenamefont {Raabe}, \citenamefont {Chang},
  \citenamefont {Zhou}, \citenamefont {Zhao}, \citenamefont {Kang},
  \citenamefont {Ju},\ and\ \citenamefont {Woo}}]{Song2020}%
  \BibitemOpen
  \bibfield  {author} {\bibinfo {author} {\bibfnamefont {Kyung~Mee}\
  \bibnamefont {Song}}, \bibinfo {author} {\bibfnamefont {Jae-Seung}\
  \bibnamefont {Jeong}}, \bibinfo {author} {\bibfnamefont {Biao}\ \bibnamefont
  {Pan}}, \bibinfo {author} {\bibfnamefont {Xichao}\ \bibnamefont {Zhang}},
  \bibinfo {author} {\bibfnamefont {Jing}\ \bibnamefont {Xia}}, \bibinfo
  {author} {\bibfnamefont {Sunkyung}\ \bibnamefont {Cha}}, \bibinfo {author}
  {\bibfnamefont {Tae-Eon}\ \bibnamefont {Park}}, \bibinfo {author}
  {\bibfnamefont {Kwangsu}\ \bibnamefont {Kim}}, \bibinfo {author}
  {\bibfnamefont {Simone}\ \bibnamefont {Finizio}}, \bibinfo {author}
  {\bibfnamefont {J{\"o}rg}\ \bibnamefont {Raabe}}, \bibinfo {author}
  {\bibfnamefont {Joonyeon}\ \bibnamefont {Chang}}, \bibinfo {author}
  {\bibfnamefont {Yan}\ \bibnamefont {Zhou}}, \bibinfo {author} {\bibfnamefont
  {Weisheng}\ \bibnamefont {Zhao}}, \bibinfo {author} {\bibfnamefont {Wang}\
  \bibnamefont {Kang}}, \bibinfo {author} {\bibfnamefont {Hyunsu}\ \bibnamefont
  {Ju}}, \ and\ \bibinfo {author} {\bibfnamefont {Seonghoon}\ \bibnamefont
  {Woo}},\ }\bibfield  {title} {\enquote {\bibinfo {title} {Skyrmion-based
  artificial synapses for neuromorphic computing},}\ }\href {\doibase
  10.1038/s41928-020-0385-0} {\bibfield  {journal} {\bibinfo  {journal} {Nature
  Electronics}\ }\textbf {\bibinfo {volume} {3}},\ \bibinfo {pages} {148--155}
  (\bibinfo {year} {2020})}\BibitemShut {NoStop}%
\bibitem [{\citenamefont {Li}\ \emph {et~al.}(2017)\citenamefont {Li},
  \citenamefont {Kang}, \citenamefont {Huang}, \citenamefont {Zhang},
  \citenamefont {Zhou},\ and\ \citenamefont {Zhao}}]{Li2017}%
  \BibitemOpen
  \bibfield  {author} {\bibinfo {author} {\bibfnamefont {Sai}\ \bibnamefont
  {Li}}, \bibinfo {author} {\bibfnamefont {Wang}\ \bibnamefont {Kang}},
  \bibinfo {author} {\bibfnamefont {Yangqi}\ \bibnamefont {Huang}}, \bibinfo
  {author} {\bibfnamefont {Xichao}\ \bibnamefont {Zhang}}, \bibinfo {author}
  {\bibfnamefont {Yan}\ \bibnamefont {Zhou}}, \ and\ \bibinfo {author}
  {\bibfnamefont {Weisheng}\ \bibnamefont {Zhao}},\ }\bibfield  {title}
  {\enquote {\bibinfo {title} {Magnetic skyrmion-based artificial neuron
  device},}\ }\href {\doibase 10.1088/1361-6528/aa7af5} {\bibfield  {journal}
  {\bibinfo  {journal} {Nanotechnology}\ }\textbf {\bibinfo {volume} {28}},\
  \bibinfo {pages} {31LT01} (\bibinfo {year} {2017})}\BibitemShut {NoStop}%
\bibitem [{\citenamefont {Chen}\ \emph {et~al.}(2018)\citenamefont {Chen},
  \citenamefont {Kang}, \citenamefont {Zhu}, \citenamefont {Zhang},
  \citenamefont {Lei}, \citenamefont {Zhang}, \citenamefont {Zhou},\ and\
  \citenamefont {Zhao}}]{Chen2018}%
  \BibitemOpen
  \bibfield  {author} {\bibinfo {author} {\bibfnamefont {Xing}\ \bibnamefont
  {Chen}}, \bibinfo {author} {\bibfnamefont {Wang}\ \bibnamefont {Kang}},
  \bibinfo {author} {\bibfnamefont {Daoqian}\ \bibnamefont {Zhu}}, \bibinfo
  {author} {\bibfnamefont {Xichao}\ \bibnamefont {Zhang}}, \bibinfo {author}
  {\bibfnamefont {Na}~\bibnamefont {Lei}}, \bibinfo {author} {\bibfnamefont
  {Youguang}\ \bibnamefont {Zhang}}, \bibinfo {author} {\bibfnamefont {Yan}\
  \bibnamefont {Zhou}}, \ and\ \bibinfo {author} {\bibfnamefont {Weisheng}\
  \bibnamefont {Zhao}},\ }\bibfield  {title} {\enquote {\bibinfo {title} {A
  compact skyrmionic leaky–integrate–fire spiking neuron device},}\ }\href
  {\doibase 10.1039/C7NR09722K} {\bibfield  {journal} {\bibinfo  {journal}
  {Nanoscale}\ }\textbf {\bibinfo {volume} {10}},\ \bibinfo {pages}
  {6139--6146} (\bibinfo {year} {2018})}\BibitemShut {NoStop}%
\bibitem [{\citenamefont {Prychynenko}\ \emph {et~al.}(2018)\citenamefont
  {Prychynenko}, \citenamefont {Sitte}, \citenamefont {Litzius}, \citenamefont
  {Kr\"uger}, \citenamefont {Bourianoff}, \citenamefont {Kl\"aui},
  \citenamefont {Sinova},\ and\ \citenamefont
  {Everschor-Sitte}}]{Prychynenko2018}%
  \BibitemOpen
  \bibfield  {author} {\bibinfo {author} {\bibfnamefont {Diana}\ \bibnamefont
  {Prychynenko}}, \bibinfo {author} {\bibfnamefont {Matthias}\ \bibnamefont
  {Sitte}}, \bibinfo {author} {\bibfnamefont {Kai}\ \bibnamefont {Litzius}},
  \bibinfo {author} {\bibfnamefont {Benjamin}\ \bibnamefont {Kr\"uger}},
  \bibinfo {author} {\bibfnamefont {George}\ \bibnamefont {Bourianoff}},
  \bibinfo {author} {\bibfnamefont {Mathias}\ \bibnamefont {Kl\"aui}}, \bibinfo
  {author} {\bibfnamefont {Jairo}\ \bibnamefont {Sinova}}, \ and\ \bibinfo
  {author} {\bibfnamefont {Karin}\ \bibnamefont {Everschor-Sitte}},\ }\bibfield
   {title} {\enquote {\bibinfo {title} {Magnetic skyrmion as a nonlinear
  resistive element: A potential building block for reservoir computing},}\
  }\href {\doibase 10.1103/PhysRevApplied.9.014034} {\bibfield  {journal}
  {\bibinfo  {journal} {Phys. Rev. Applied}\ }\textbf {\bibinfo {volume} {9}},\
  \bibinfo {pages} {014034} (\bibinfo {year} {2018})}\BibitemShut {NoStop}%
\bibitem [{\citenamefont {Bourianoff}\ \emph {et~al.}(2018)\citenamefont
  {Bourianoff}, \citenamefont {Pinna}, \citenamefont {Sitte},\ and\
  \citenamefont {Everschor-Sitte}}]{Bourianoff2018}%
  \BibitemOpen
  \bibfield  {author} {\bibinfo {author} {\bibfnamefont {George}\ \bibnamefont
  {Bourianoff}}, \bibinfo {author} {\bibfnamefont {Daniele}\ \bibnamefont
  {Pinna}}, \bibinfo {author} {\bibfnamefont {Matthias}\ \bibnamefont {Sitte}},
  \ and\ \bibinfo {author} {\bibfnamefont {Karin}\ \bibnamefont
  {Everschor-Sitte}},\ }\bibfield  {title} {\enquote {\bibinfo {title}
  {Potential implementation of reservoir computing models based on magnetic
  skyrmions},}\ }\href {\doibase 10.1063/1.5006918} {\bibfield  {journal}
  {\bibinfo  {journal} {AIP Advances}\ }\textbf {\bibinfo {volume} {8}},\
  \bibinfo {pages} {055602} (\bibinfo {year} {2018})},\ \Eprint
  {http://arxiv.org/abs/https://doi.org/10.1063/1.5006918}
  {https://doi.org/10.1063/1.5006918} \BibitemShut {NoStop}%
\bibitem [{\citenamefont {Pinna}\ \emph {et~al.}(2018)\citenamefont {Pinna},
  \citenamefont {Abreu~Araujo}, \citenamefont {Kim}, \citenamefont {Cros},
  \citenamefont {Querlioz}, \citenamefont {Bessiere}, \citenamefont {Droulez},\
  and\ \citenamefont {Grollier}}]{Pinna2018}%
  \BibitemOpen
  \bibfield  {author} {\bibinfo {author} {\bibfnamefont {D.}~\bibnamefont
  {Pinna}}, \bibinfo {author} {\bibfnamefont {F.}~\bibnamefont {Abreu~Araujo}},
  \bibinfo {author} {\bibfnamefont {J.-V.}\ \bibnamefont {Kim}}, \bibinfo
  {author} {\bibfnamefont {V.}~\bibnamefont {Cros}}, \bibinfo {author}
  {\bibfnamefont {D.}~\bibnamefont {Querlioz}}, \bibinfo {author}
  {\bibfnamefont {P.}~\bibnamefont {Bessiere}}, \bibinfo {author}
  {\bibfnamefont {J.}~\bibnamefont {Droulez}}, \ and\ \bibinfo {author}
  {\bibfnamefont {J.}~\bibnamefont {Grollier}},\ }\bibfield  {title} {\enquote
  {\bibinfo {title} {Skyrmion gas manipulation for probabilistic computing},}\
  }\href {\doibase 10.1103/PhysRevApplied.9.064018} {\bibfield  {journal}
  {\bibinfo  {journal} {Phys. Rev. Applied}\ }\textbf {\bibinfo {volume} {9}},\
  \bibinfo {pages} {064018} (\bibinfo {year} {2018})}\BibitemShut {NoStop}%
\bibitem [{\citenamefont {Z{\'a}zvorka}\ \emph {et~al.}(2019)\citenamefont
  {Z{\'a}zvorka}, \citenamefont {Jakobs}, \citenamefont {Heinze}, \citenamefont
  {Keil}, \citenamefont {Kromin}, \citenamefont {Jaiswal}, \citenamefont
  {Litzius}, \citenamefont {Jakob}, \citenamefont {Virnau}, \citenamefont
  {Pinna}, \citenamefont {Everschor-Sitte}, \citenamefont {R{\'o}zsa},
  \citenamefont {Donges}, \citenamefont {Nowak},\ and\ \citenamefont
  {Kl{\"a}ui}}]{Zazvorka2019}%
  \BibitemOpen
  \bibfield  {author} {\bibinfo {author} {\bibfnamefont {Jakub}\ \bibnamefont
  {Z{\'a}zvorka}}, \bibinfo {author} {\bibfnamefont {Florian}\ \bibnamefont
  {Jakobs}}, \bibinfo {author} {\bibfnamefont {Daniel}\ \bibnamefont {Heinze}},
  \bibinfo {author} {\bibfnamefont {Niklas}\ \bibnamefont {Keil}}, \bibinfo
  {author} {\bibfnamefont {Sascha}\ \bibnamefont {Kromin}}, \bibinfo {author}
  {\bibfnamefont {Samridh}\ \bibnamefont {Jaiswal}}, \bibinfo {author}
  {\bibfnamefont {Kai}\ \bibnamefont {Litzius}}, \bibinfo {author}
  {\bibfnamefont {Gerhard}\ \bibnamefont {Jakob}}, \bibinfo {author}
  {\bibfnamefont {Peter}\ \bibnamefont {Virnau}}, \bibinfo {author}
  {\bibfnamefont {Daniele}\ \bibnamefont {Pinna}}, \bibinfo {author}
  {\bibfnamefont {Karin}\ \bibnamefont {Everschor-Sitte}}, \bibinfo {author}
  {\bibfnamefont {Levente}\ \bibnamefont {R{\'o}zsa}}, \bibinfo {author}
  {\bibfnamefont {Andreas}\ \bibnamefont {Donges}}, \bibinfo {author}
  {\bibfnamefont {Ulrich}\ \bibnamefont {Nowak}}, \ and\ \bibinfo {author}
  {\bibfnamefont {Mathias}\ \bibnamefont {Kl{\"a}ui}},\ }\bibfield  {title}
  {\enquote {\bibinfo {title} {Thermal skyrmion diffusion used in a reshuffler
  device},}\ }\href {\doibase 10.1038/s41565-019-0436-8} {\bibfield  {journal}
  {\bibinfo  {journal} {Nature Nanotechnology}\ }\textbf {\bibinfo {volume}
  {14}},\ \bibinfo {pages} {658--661} (\bibinfo {year} {2019})}\BibitemShut
  {NoStop}%
\bibitem [{\citenamefont {{Yao}}\ \emph {et~al.}(2020)\citenamefont {{Yao}},
  \citenamefont {{Chen}}, \citenamefont {{Kang}}, \citenamefont {{Zhang}},\
  and\ \citenamefont {{Zhao}}}]{Yao2020}%
  \BibitemOpen
  \bibfield  {author} {\bibinfo {author} {\bibfnamefont {Y.}~\bibnamefont
  {{Yao}}}, \bibinfo {author} {\bibfnamefont {X.}~\bibnamefont {{Chen}}},
  \bibinfo {author} {\bibfnamefont {W.}~\bibnamefont {{Kang}}}, \bibinfo
  {author} {\bibfnamefont {Y.}~\bibnamefont {{Zhang}}}, \ and\ \bibinfo
  {author} {\bibfnamefont {W.}~\bibnamefont {{Zhao}}},\ }\bibfield  {title}
  {\enquote {\bibinfo {title} {Thermal brownian motion of skyrmion for true
  random number generation},}\ }\href {\doibase 10.1109/TED.2020.2989420}
  {\bibfield  {journal} {\bibinfo  {journal} {IEEE Transactions on Electron
  Devices}\ }\textbf {\bibinfo {volume} {67}},\ \bibinfo {pages} {2553--2558}
  (\bibinfo {year} {2020})}\BibitemShut {NoStop}%
\bibitem [{\citenamefont {Zhang}\ \emph
  {et~al.}(2015{\natexlab{b}})\citenamefont {Zhang}, \citenamefont {Wang},
  \citenamefont {Zheng}, \citenamefont {Zhu}, \citenamefont {Liu},
  \citenamefont {Chen}, \citenamefont {Jin}, \citenamefont {Liu}, \citenamefont
  {Jia},\ and\ \citenamefont {Xue}}]{Zhang2015b}%
  \BibitemOpen
  \bibfield  {author} {\bibinfo {author} {\bibfnamefont {Senfu}\ \bibnamefont
  {Zhang}}, \bibinfo {author} {\bibfnamefont {Jianbo}\ \bibnamefont {Wang}},
  \bibinfo {author} {\bibfnamefont {Qi}~\bibnamefont {Zheng}}, \bibinfo
  {author} {\bibfnamefont {Qiyuan}\ \bibnamefont {Zhu}}, \bibinfo {author}
  {\bibfnamefont {Xianyin}\ \bibnamefont {Liu}}, \bibinfo {author}
  {\bibfnamefont {Shujun}\ \bibnamefont {Chen}}, \bibinfo {author}
  {\bibfnamefont {Chendong}\ \bibnamefont {Jin}}, \bibinfo {author}
  {\bibfnamefont {Qingfang}\ \bibnamefont {Liu}}, \bibinfo {author}
  {\bibfnamefont {Chenglong}\ \bibnamefont {Jia}}, \ and\ \bibinfo {author}
  {\bibfnamefont {Desheng}\ \bibnamefont {Xue}},\ }\bibfield  {title} {\enquote
  {\bibinfo {title} {Current-induced magnetic skyrmions oscillator},}\ }\href
  {\doibase 10.1088/1367-2630/17/2/023061} {\bibfield  {journal} {\bibinfo
  {journal} {New Journal of Physics}\ }\textbf {\bibinfo {volume} {17}},\
  \bibinfo {pages} {023061} (\bibinfo {year} {2015}{\natexlab{b}})}\BibitemShut
  {NoStop}%
\bibitem [{\citenamefont {Finocchio}\ \emph {et~al.}(2015)\citenamefont
  {Finocchio}, \citenamefont {Ricci}, \citenamefont {Tomasello}, \citenamefont
  {Giordano}, \citenamefont {Lanuzza}, \citenamefont {Puliafito}, \citenamefont
  {Burrascano}, \citenamefont {Azzerboni},\ and\ \citenamefont
  {Carpentieri}}]{Finocchio2015}%
  \BibitemOpen
  \bibfield  {author} {\bibinfo {author} {\bibfnamefont {G.}~\bibnamefont
  {Finocchio}}, \bibinfo {author} {\bibfnamefont {M.}~\bibnamefont {Ricci}},
  \bibinfo {author} {\bibfnamefont {R.}~\bibnamefont {Tomasello}}, \bibinfo
  {author} {\bibfnamefont {A.}~\bibnamefont {Giordano}}, \bibinfo {author}
  {\bibfnamefont {M.}~\bibnamefont {Lanuzza}}, \bibinfo {author} {\bibfnamefont
  {V.}~\bibnamefont {Puliafito}}, \bibinfo {author} {\bibfnamefont
  {P.}~\bibnamefont {Burrascano}}, \bibinfo {author} {\bibfnamefont
  {B.}~\bibnamefont {Azzerboni}}, \ and\ \bibinfo {author} {\bibfnamefont
  {M.}~\bibnamefont {Carpentieri}},\ }\bibfield  {title} {\enquote {\bibinfo
  {title} {Skyrmion based microwave detectors and harvesting},}\ }\href
  {\doibase 10.1063/1.4938539} {\bibfield  {journal} {\bibinfo  {journal}
  {Applied Physics Letters}\ }\textbf {\bibinfo {volume} {107}},\ \bibinfo
  {pages} {262401} (\bibinfo {year} {2015})},\ \Eprint
  {http://arxiv.org/abs/https://doi.org/10.1063/1.4938539}
  {https://doi.org/10.1063/1.4938539} \BibitemShut {NoStop}%
\bibitem [{\citenamefont {Str\'o\ifmmode~\dot{z}\else \.{z}\fi{}ecka}\ \emph
  {et~al.}(2011)\citenamefont {Str\'o\ifmmode~\dot{z}\else \.{z}\fi{}ecka},
  \citenamefont {Eiguren},\ and\ \citenamefont {Pascual}}]{Strozecka:2011}%
  \BibitemOpen
  \bibfield  {author} {\bibinfo {author} {\bibfnamefont {Anna}\ \bibnamefont
  {Str\'o\ifmmode~\dot{z}\else \.{z}\fi{}ecka}}, \bibinfo {author}
  {\bibfnamefont {Asier}\ \bibnamefont {Eiguren}}, \ and\ \bibinfo {author}
  {\bibfnamefont {Jose~Ignacio}\ \bibnamefont {Pascual}},\ }\bibfield  {title}
  {\enquote {\bibinfo {title} {Quasiparticle interference around a magnetic
  impurity on a surface with strong spin-orbit coupling},}\ }\href {\doibase
  10.1103/PhysRevLett.107.186805} {\bibfield  {journal} {\bibinfo  {journal}
  {Phys. Rev. Lett.}\ }\textbf {\bibinfo {volume} {107}},\ \bibinfo {pages}
  {186805} (\bibinfo {year} {2011})}\BibitemShut {NoStop}%
\bibitem [{\citenamefont {Denisov}\ \emph {et~al.}(2019)\citenamefont
  {Denisov}, \citenamefont {Rozhansky}, \citenamefont {Averkiev},\ and\
  \citenamefont {L{\"a}hderanta}}]{Denisov:2019}%
  \BibitemOpen
  \bibfield  {author} {\bibinfo {author} {\bibfnamefont {K.~S.}\ \bibnamefont
  {Denisov}}, \bibinfo {author} {\bibfnamefont {I.~V.}\ \bibnamefont
  {Rozhansky}}, \bibinfo {author} {\bibfnamefont {N.~S.}\ \bibnamefont
  {Averkiev}}, \ and\ \bibinfo {author} {\bibfnamefont {E.}~\bibnamefont
  {L{\"a}hderanta}},\ }\bibfield  {title} {\enquote {\bibinfo {title} {Chiral
  spin ordering of electron gas in solids with broken time reversal
  symmetry},}\ }\href {\doibase 10.1038/s41598-019-47274-6} {\bibfield
  {journal} {\bibinfo  {journal} {Scientific Reports}\ }\textbf {\bibinfo
  {volume} {9}},\ \bibinfo {pages} {10817} (\bibinfo {year}
  {2019})}\BibitemShut {NoStop}%
\bibitem [{\citenamefont {Wang}\ \emph {et~al.}(2020)\citenamefont {Wang},
  \citenamefont {Xu},\ and\ \citenamefont {Lai}}]{Wang:2020}%
  \BibitemOpen
  \bibfield  {author} {\bibinfo {author} {\bibfnamefont {Cheng-Zhen}\
  \bibnamefont {Wang}}, \bibinfo {author} {\bibfnamefont {Hong-Ya}\
  \bibnamefont {Xu}}, \ and\ \bibinfo {author} {\bibfnamefont {Ying-Cheng}\
  \bibnamefont {Lai}},\ }\bibfield  {title} {\enquote {\bibinfo {title}
  {Scattering of dirac electrons from a skyrmion: Emergence of robust skew
  scattering},}\ }\href {\doibase 10.1103/PhysRevResearch.2.013247} {\bibfield
  {journal} {\bibinfo  {journal} {Phys. Rev. Research}\ }\textbf {\bibinfo
  {volume} {2}},\ \bibinfo {pages} {013247} (\bibinfo {year}
  {2020})}\BibitemShut {NoStop}%
\bibitem [{\citenamefont {Romming}\ \emph {et~al.}(2013)\citenamefont
  {Romming}, \citenamefont {Hanneken}, \citenamefont {Menzel}, \citenamefont
  {Bickel}, \citenamefont {Wolter}, \citenamefont {von Bergmann}, \citenamefont
  {Kubetzka},\ and\ \citenamefont {Wiesendanger}}]{Romming2013}%
  \BibitemOpen
  \bibfield  {author} {\bibinfo {author} {\bibfnamefont {Niklas}\ \bibnamefont
  {Romming}}, \bibinfo {author} {\bibfnamefont {Christian}\ \bibnamefont
  {Hanneken}}, \bibinfo {author} {\bibfnamefont {Matthias}\ \bibnamefont
  {Menzel}}, \bibinfo {author} {\bibfnamefont {Jessica~E.}\ \bibnamefont
  {Bickel}}, \bibinfo {author} {\bibfnamefont {Boris}\ \bibnamefont {Wolter}},
  \bibinfo {author} {\bibfnamefont {Kirsten}\ \bibnamefont {von Bergmann}},
  \bibinfo {author} {\bibfnamefont {Andr{\'e}}\ \bibnamefont {Kubetzka}}, \
  and\ \bibinfo {author} {\bibfnamefont {Roland}\ \bibnamefont
  {Wiesendanger}},\ }\bibfield  {title} {\enquote {\bibinfo {title} {Writing
  and deleting single magnetic skyrmions},}\ }\href {\doibase
  10.1126/science.1240573} {\bibfield  {journal} {\bibinfo  {journal}
  {Science}\ }\textbf {\bibinfo {volume} {341}},\ \bibinfo {pages} {636--639}
  (\bibinfo {year} {2013})}\BibitemShut {NoStop}%
\bibitem [{\citenamefont {Romming}\ \emph {et~al.}(2015)\citenamefont
  {Romming}, \citenamefont {Kubetzka}, \citenamefont {Hanneken}, \citenamefont
  {von Bergmann},\ and\ \citenamefont {Wiesendanger}}]{Romming2015}%
  \BibitemOpen
  \bibfield  {author} {\bibinfo {author} {\bibfnamefont {Niklas}\ \bibnamefont
  {Romming}}, \bibinfo {author} {\bibfnamefont {Andr\'e}\ \bibnamefont
  {Kubetzka}}, \bibinfo {author} {\bibfnamefont {Christian}\ \bibnamefont
  {Hanneken}}, \bibinfo {author} {\bibfnamefont {Kirsten}\ \bibnamefont {von
  Bergmann}}, \ and\ \bibinfo {author} {\bibfnamefont {Roland}\ \bibnamefont
  {Wiesendanger}},\ }\bibfield  {title} {\enquote {\bibinfo {title}
  {Field-dependent size and shape of single magnetic skyrmions},}\ }\href@noop
  {} {\bibfield  {journal} {\bibinfo  {journal} {Phys. Rev. Lett.}\ }\textbf
  {\bibinfo {volume} {114}},\ \bibinfo {pages} {177203} (\bibinfo {year}
  {2015})}\BibitemShut {NoStop}%
\bibitem [{\citenamefont {Dzyaloshinsky}(1958)}]{Dzyalosinkii}%
  \BibitemOpen
  \bibfield  {author} {\bibinfo {author} {\bibfnamefont {I.}~\bibnamefont
  {Dzyaloshinsky}},\ }\bibfield  {title} {\enquote {\bibinfo {title} {A
  thermodynamic theory of “weak” ferromagnetism of antiferromagnetics},}\
  }\href@noop {} {\bibfield  {journal} {\bibinfo  {journal} {J. Phys. Chem.
  Solid}\ }\textbf {\bibinfo {volume} {4(4)}},\ \bibinfo {pages} {241–255}
  (\bibinfo {year} {1958})}\BibitemShut {NoStop}%
\bibitem [{\citenamefont {Moriya}(1960)}]{Moriya}%
  \BibitemOpen
  \bibfield  {author} {\bibinfo {author} {\bibfnamefont {T.}~\bibnamefont
  {Moriya}},\ }\bibfield  {title} {\enquote {\bibinfo {title} {Anisotropic
  superexchange interaction and weak ferromagnetism},}\ }\href@noop {}
  {\bibfield  {journal} {\bibinfo  {journal} {Phys. Rev.}\ }\textbf {\bibinfo
  {volume} {120}},\ \bibinfo {pages} {91} (\bibinfo {year} {1960})}\BibitemShut
  {NoStop}%
\bibitem [{\citenamefont {Bouhassoune}\ \emph {et~al.}(2019)\citenamefont
  {Bouhassoune}, \citenamefont {Fernandes}, \citenamefont {Blügel},\ and\
  \citenamefont {Lounis}}]{Bouhassoune2019}%
  \BibitemOpen
  \bibfield  {author} {\bibinfo {author} {\bibfnamefont {Mohammed}\
  \bibnamefont {Bouhassoune}}, \bibinfo {author} {\bibfnamefont {Imara~Lima}\
  \bibnamefont {Fernandes}}, \bibinfo {author} {\bibfnamefont {Stefan}\
  \bibnamefont {Blügel}}, \ and\ \bibinfo {author} {\bibfnamefont {Samir}\
  \bibnamefont {Lounis}},\ }\bibfield  {title} {\enquote {\bibinfo {title}
  {Unoccupied surface and interface states in pd thin films deposited on
  fe/ir(111) surface},}\ }\href {\doibase 10.1088/1367-2630/ab2487} {\bibfield
  {journal} {\bibinfo  {journal} {New Journal of Physics}\ }\textbf {\bibinfo
  {volume} {21}},\ \bibinfo {pages} {063015} (\bibinfo {year}
  {2019})}\BibitemShut {NoStop}%
\bibitem [{\citenamefont {Hanneken}\ \emph {et~al.}(2015)\citenamefont
  {Hanneken}, \citenamefont {Otte}, \citenamefont {Kubetzka}, \citenamefont
  {Dup{\'e}}, \citenamefont {Romming}, \citenamefont {Von~Bergmann},
  \citenamefont {Wiesendanger},\ and\ \citenamefont {Heinze}}]{Hanneken2015}%
  \BibitemOpen
  \bibfield  {author} {\bibinfo {author} {\bibfnamefont {Christian}\
  \bibnamefont {Hanneken}}, \bibinfo {author} {\bibfnamefont {Fabian}\
  \bibnamefont {Otte}}, \bibinfo {author} {\bibfnamefont {Andr{\'e}}\
  \bibnamefont {Kubetzka}}, \bibinfo {author} {\bibfnamefont {Bertrand}\
  \bibnamefont {Dup{\'e}}}, \bibinfo {author} {\bibfnamefont {Niklas}\
  \bibnamefont {Romming}}, \bibinfo {author} {\bibfnamefont {Kirsten}\
  \bibnamefont {Von~Bergmann}}, \bibinfo {author} {\bibfnamefont {Roland}\
  \bibnamefont {Wiesendanger}}, \ and\ \bibinfo {author} {\bibfnamefont
  {Stefan}\ \bibnamefont {Heinze}},\ }\bibfield  {title} {\enquote {\bibinfo
  {title} {Electrical detection of magnetic skyrmions by tunnelling
  non-collinear magnetoresistance},}\ }\href@noop {} {\bibfield  {journal}
  {\bibinfo  {journal} {Nat. Nanotechnol.}\ }\textbf {\bibinfo {volume} {10}},\
  \bibinfo {pages} {1039} (\bibinfo {year} {2015})}\BibitemShut {NoStop}%
\bibitem [{\citenamefont {Bauer}(2013)}]{Bauer2013}%
  \BibitemOpen
  \bibfield  {author} {\bibinfo {author} {\bibfnamefont {D.~S.~G.}\
  \bibnamefont {Bauer}},\ }\bibfield  {title} {\enquote {\bibinfo {title}
  {Development of a relativistic full-potential first-principles multiple
  scattering green function method applied to complex magnetic textures of nano
  structures at surfaces},}\ }\href@noop {} {\bibfield  {journal} {\bibinfo
  {journal} {PhD dissertation at the RWTH-Aachen}\ } (\bibinfo {year}
  {2013})}\BibitemShut {NoStop}%
\bibitem [{\citenamefont {Papanikolaou}\ \emph {et~al.}(2002)\citenamefont
  {Papanikolaou}, \citenamefont {Zeller},\ and\ \citenamefont
  {Dederichs}}]{Papanikolaou2002}%
  \BibitemOpen
  \bibfield  {author} {\bibinfo {author} {\bibfnamefont {N}~\bibnamefont
  {Papanikolaou}}, \bibinfo {author} {\bibfnamefont {R}~\bibnamefont {Zeller}},
  \ and\ \bibinfo {author} {\bibfnamefont {P~H}\ \bibnamefont {Dederichs}},\
  }\bibfield  {title} {\enquote {\bibinfo {title} {{Conceptual improvements of
  the KKR method}},}\ }\href@noop {} {\bibfield  {journal} {\bibinfo  {journal}
  {Journal of Physics: Condensed Matter}\ }\textbf {\bibinfo {volume} {14}},\
  \bibinfo {pages} {2799} (\bibinfo {year} {2002})}\BibitemShut {NoStop}%
\bibitem [{\citenamefont {Vosko}\ \emph {et~al.}(1980)\citenamefont {Vosko},
  \citenamefont {Wilk},\ and\ \citenamefont {Nusair}}]{Vosko1980}%
  \BibitemOpen
  \bibfield  {author} {\bibinfo {author} {\bibfnamefont {Seymour~H}\
  \bibnamefont {Vosko}}, \bibinfo {author} {\bibfnamefont {Leslie}\
  \bibnamefont {Wilk}}, \ and\ \bibinfo {author} {\bibfnamefont {Marwan}\
  \bibnamefont {Nusair}},\ }\bibfield  {title} {\enquote {\bibinfo {title}
  {Accurate spin-dependent electron liquid correlation energies for local spin
  density calculations: a critical analysis},}\ }\href@noop {} {\bibfield
  {journal} {\bibinfo  {journal} {Canadian Journal of physics}\ }\textbf
  {\bibinfo {volume} {58}},\ \bibinfo {pages} {1200--1211} (\bibinfo {year}
  {1980})}\BibitemShut {NoStop}%
\bibitem [{\citenamefont {Dup{\'e}}\ \emph {et~al.}(2014)\citenamefont
  {Dup{\'e}}, \citenamefont {Hoffmann}, \citenamefont {Paillard},\ and\
  \citenamefont {Heinze}}]{Dupe2014}%
  \BibitemOpen
  \bibfield  {author} {\bibinfo {author} {\bibfnamefont {B.}~\bibnamefont
  {Dup{\'e}}}, \bibinfo {author} {\bibfnamefont {M.}~\bibnamefont {Hoffmann}},
  \bibinfo {author} {\bibfnamefont {C.}~\bibnamefont {Paillard}}, \ and\
  \bibinfo {author} {\bibfnamefont {S.}~\bibnamefont {Heinze}},\ }\bibfield
  {title} {\enquote {\bibinfo {title} {Tailoring magnetic skyrmions in
  ultra-thin transition metal films},}\ }\href@noop {} {\bibfield  {journal}
  {\bibinfo  {journal} {Nature Commun.}\ }\textbf {\bibinfo {volume} {5}},\
  \bibinfo {pages} {4030} (\bibinfo {year} {2014})}\BibitemShut {NoStop}%
\bibitem [{\citenamefont {Garcia-Moliner}\ and\ \citenamefont
  {Velasco}(1986)}]{Moliner1986}%
  \BibitemOpen
  \bibfield  {author} {\bibinfo {author} {\bibfnamefont {F.}~\bibnamefont
  {Garcia-Moliner}}\ and\ \bibinfo {author} {\bibfnamefont {V.R.}\ \bibnamefont
  {Velasco}},\ }\bibfield  {title} {\enquote {\bibinfo {title} {Theory of
  incomplete crystals, surfaces, defects, interfaces and layered structures},}\
  }\href {\doibase https://doi.org/10.1016/0079-6816(86)90011-0} {\bibfield
  {journal} {\bibinfo  {journal} {Progress in Surface Science}\ }\textbf
  {\bibinfo {volume} {21}},\ \bibinfo {pages} {93 -- 162} (\bibinfo {year}
  {1986})}\BibitemShut {NoStop}%
\bibitem [{\citenamefont {Szunyogh}\ \emph {et~al.}(1994)\citenamefont
  {Szunyogh}, \citenamefont {\'Ujfalussy}, \citenamefont {Weinberger},\ and\
  \citenamefont {Koll\'ar}}]{Szunyogh1994}%
  \BibitemOpen
  \bibfield  {author} {\bibinfo {author} {\bibfnamefont {L.}~\bibnamefont
  {Szunyogh}}, \bibinfo {author} {\bibfnamefont {B.}~\bibnamefont
  {\'Ujfalussy}}, \bibinfo {author} {\bibfnamefont {P.}~\bibnamefont
  {Weinberger}}, \ and\ \bibinfo {author} {\bibfnamefont {J.}~\bibnamefont
  {Koll\'ar}},\ }\bibfield  {title} {\enquote {\bibinfo {title}
  {Self-consistent localized kkr scheme for surfaces and interfaces},}\ }\href
  {\doibase 10.1103/PhysRevB.49.2721} {\bibfield  {journal} {\bibinfo
  {journal} {Phys. Rev. B}\ }\textbf {\bibinfo {volume} {49}},\ \bibinfo
  {pages} {2721--2729} (\bibinfo {year} {1994})}\BibitemShut {NoStop}%
\bibitem [{\citenamefont {Tersoff}\ and\ \citenamefont
  {Hamann}(1983)}]{Tersoff1983}%
  \BibitemOpen
  \bibfield  {author} {\bibinfo {author} {\bibfnamefont {J.}~\bibnamefont
  {Tersoff}}\ and\ \bibinfo {author} {\bibfnamefont {D.~R.}\ \bibnamefont
  {Hamann}},\ }\bibfield  {title} {\enquote {\bibinfo {title} {{Theory and
  Application for the Scanning Tunneling Microscope}},}\ }\href {\doibase
  10.1103/PhysRevLett.50.1998} {\bibfield  {journal} {\bibinfo  {journal}
  {Phys. Rev. Lett.}\ }\textbf {\bibinfo {volume} {50}},\ \bibinfo {pages}
  {1998--2001} (\bibinfo {year} {1983})}\BibitemShut {NoStop}%
\bibitem [{\citenamefont {Palotás}\ \emph {et~al.}(2021)\citenamefont
  {Palotás}, \citenamefont {Rózsa}, \citenamefont {Simon},\ and\
  \citenamefont {Szunyogh}}]{Palotas2021}%
  \BibitemOpen
  \bibfield  {author} {\bibinfo {author} {\bibfnamefont {Krisztián}\
  \bibnamefont {Palotás}}, \bibinfo {author} {\bibfnamefont {Levente}\
  \bibnamefont {Rózsa}}, \bibinfo {author} {\bibfnamefont {Eszter}\
  \bibnamefont {Simon}}, \ and\ \bibinfo {author} {\bibfnamefont {László}\
  \bibnamefont {Szunyogh}},\ }\bibfield  {title} {\enquote {\bibinfo {title}
  {High-resolution tunneling spin transport characteristics of topologically
  distinct magnetic skyrmionic textures from theoretical calculations},}\
  }\href {\doibase https://doi.org/10.1016/j.jmmm.2020.167440} {\bibfield
  {journal} {\bibinfo  {journal} {Journal of Magnetism and Magnetic Materials}\
  }\textbf {\bibinfo {volume} {519}},\ \bibinfo {pages} {167440} (\bibinfo
  {year} {2021})}\BibitemShut {NoStop}%
\end{thebibliography}%

\end{document}